\pdfoutput=1
\documentclass[11pt,twoside,a4paper,cmspaper,final,collab]{cms-tdr}
\def\svnVersion{408924f}\def\svnDate{2026/05/07}\def\cmsCernNoTag{CERN-EP-2025-301}\def\cmsCernDate{\today}\def\cmsMessage{Published in Physics Letters B as \href{http://dx.doi.org/10.1016/j.physletb.2026.140498}{\doi{10.1016/j.physletb.2026.140498}.}}
\begin{document}\cmsNoteHeader{EXO-25-010}

\newlength\cmsTabSkip\setlength{\cmsTabSkip}{1ex}
\ifthenelse{\boolean{cms@external}}{
  \providecommand{\cmsLeft}{upper\xspace}
  \providecommand{\cmsRight}{lower\xspace}
  \newcommand{\cmsApp}{}
  }{
  \providecommand{\cmsRight}{right\xspace}
  \providecommand{\cmsLeft}{left\xspace}
  \newcommand{\cmsApp}{\appendixname\xspace}}

\cmsNoteHeader{EXO-25-010}
\title{Search for heavy long-lived charged particles with level-1 trigger scouting data from proton-proton collisions at \texorpdfstring{$\sqrt{s} = 13.6\TeV$}{sqrt(s) = 13.6 TeV}}

\author*[cern]{The CMS Collaboration}

\date{\today}

\abstract{
	A search for heavy long-lived charged particles at the LHC is presented. Particles traversing the CMS muon detector across several bunch crossings are searched for using a data sample of proton-proton collisions at $\sqrt{s}=13.6\TeV$ collected with the CMS detector in 2024, corresponding to an integrated luminosity of 3.7\fbinv. This is the first search using the novel level-1 trigger scouting data set collected without any trigger selection, allowing correlations between bunch crossings to be analyzed. The results are interpreted as upper limits on the cross sections of several benchmark processes with pair production of heavy long-lived charged particles. Upper limits on the fiducial cross section of a heavy long-lived charged particle with $\pt>500\GeV$ and $\abs{\eta}<0.83$ are also set in different ranges of $\beta=v/c$. This analysis is a proof of concept for the level-1 trigger data scouting system and complements existing searches for heavy long-lived charged particles by extending the sensitivity to lower $\beta$ values. 
}

\hypersetup{
pdfauthor={CMS Collaboration},
pdftitle={Search for heavy long-lived charged particles with level-1 trigger scouting data from proton-proton collisions at \texorpdfstring{$\sqrt{s} = 13.6\TeV$}{sqrt(s) = 13.6 TeV}},
pdfsubject={CMS},
pdfkeywords={CMS, HSCP, long-lived, level-1 trigger, data scouting}}

\maketitle 

\section{Introduction}

Heavy long-lived charged-particle searches are motivated by multiple beyond the standard model scenarios~\cite{Fairbairn:2006gg,Strassler:2006im,Bauer:2009cc},
such as supersymmetry,
extra dimensions, and other theoretical frameworks. 
These particles are generically 
referred to as heavy stable charged particles (HSCPs),
even if they may be metastable. 
If produced in proton-proton ($\Pp\Pp$) collisions at the CERN LHC,
HSCPs would traverse the detector leaving distinctive experimental signatures. 
Because of their large mass and long lifetime, slow-moving HSCPs 
typically exhibit
high ionization energy loss and unusually long time-of-flight.

Past collider searches for HSCPs at the CERN LEP~\cite{ALEPH:1997siq,DELPHI:2000urg,L3:2001xsz,OPAL:2003zpa,H1:2004maq}, Fermilab Tevatron~\cite{CDF:2009xuh,D0:2008jxy,D0:2011cqo, PhysRevD.87.052011}, and LHC~\cite{ATLAS:2011zgn,ATLAS:2011ghv,ATLAS:2011wuq,ATLAS:2012urj,ATLAS:2013cab,ATLAS:2014fka,ATLAS:2015hau,ATLAS:2015wsk,ATLAS:2016tbt,ATLAS:2016onr,CMS:2011arq,CMS:2012xi,CMS:2012wcg,CMS:2013czn,CMS:2016kce,ATLAS:2022pib,CMS:2024eyx,CMS:2024nhn,ATLAS:2025fdm} have set stringent constraints on various models, with 
limits on gluino $R$-hadrons, top squark $R$-hadrons, and fourth-generation leptons reaching the \TeV scale. 
The recent HSCP search by the CMS Collaboration~\cite{CMS:2024nhn}
looks for an isolated high transverse momentum (\pt) track that deposits an anomalous energy per unit length
in the silicon tracker and relies on muon-based triggers to select events online.
Because of this triggering strategy, it is sensitive to HSCPs with $\beta = v/c \gtrsim 0.6$, which can cross the CMS 
detector within a single bunch crossing (BX). To pass
 muon-based triggers, the muon tracks reconstructed in the tracker and in the muon detector must be in the same BX.
Very slow particles may take several BXs to reach and cross the muon detector, causing the trigger to fail.
While it is possible to indirectly trigger on missing transverse momentum in the calorimeters, 
such triggers typically have low and model-dependent signal efficiency. 
Even when the event can be triggered on, low-$\beta$ particles may not be fully reconstructed 
if their trajectories span several BXs. Because of this limitation, past LHC searches have only targeted HSCP masses up to 3\TeV. 

The analysis presented in this Letter avoids the triggering and reconstruction constraints by relying on the data collected without 
trigger conditions by the novel CMS level-1 trigger data scouting (L1DS) system~\cite{CERN-LHCC-2020-004,CMS:2023gfb,migliorini_2024_wvep1-7db41,ardino_2024_qhb4p-wx311}. 
It probes the existence of HSCPs with mass between 1 and 6.5\TeV leaving a signature in the muon detector before decaying.
The search is based on data collected in the second half of 2024 in $\Pp\Pp$ collisions at $\sqrt{s}=13.6\TeV$, 
corresponding to an integrated luminosity of 3.7\fbinv. 
This is the first analysis relying 
on data from the L1DS system and, as such, it is a crucial proof of concept for the physics potential of the L1DS.
Tabulated results are provided in the HEPData record for this analysis~\cite{hepdata}.

\section{The CMS detector and its hardware muon trigger}

The CMS apparatus~\cite{CMS:2008xjf,CMS:2023gfb} is a multipurpose, nearly hermetic detector. 
Its central feature is a superconducting solenoid of 6\unit{m} internal diameter, providing a magnetic field of 3.8\unit{T}. 
Events of interest are usually selected using a two-tiered trigger system, where the level-1 trigger (L1),  
composed of custom hardware processors, uses information from the calorimeters and muon detectors to select events 
at a rate of around 100\unit{kHz} within a fixed latency of 4\mus~\cite{CMS:2020cmk}. 

\begin{figure}[htbp]
  \centering
  \includegraphics[width=0.49\textwidth]{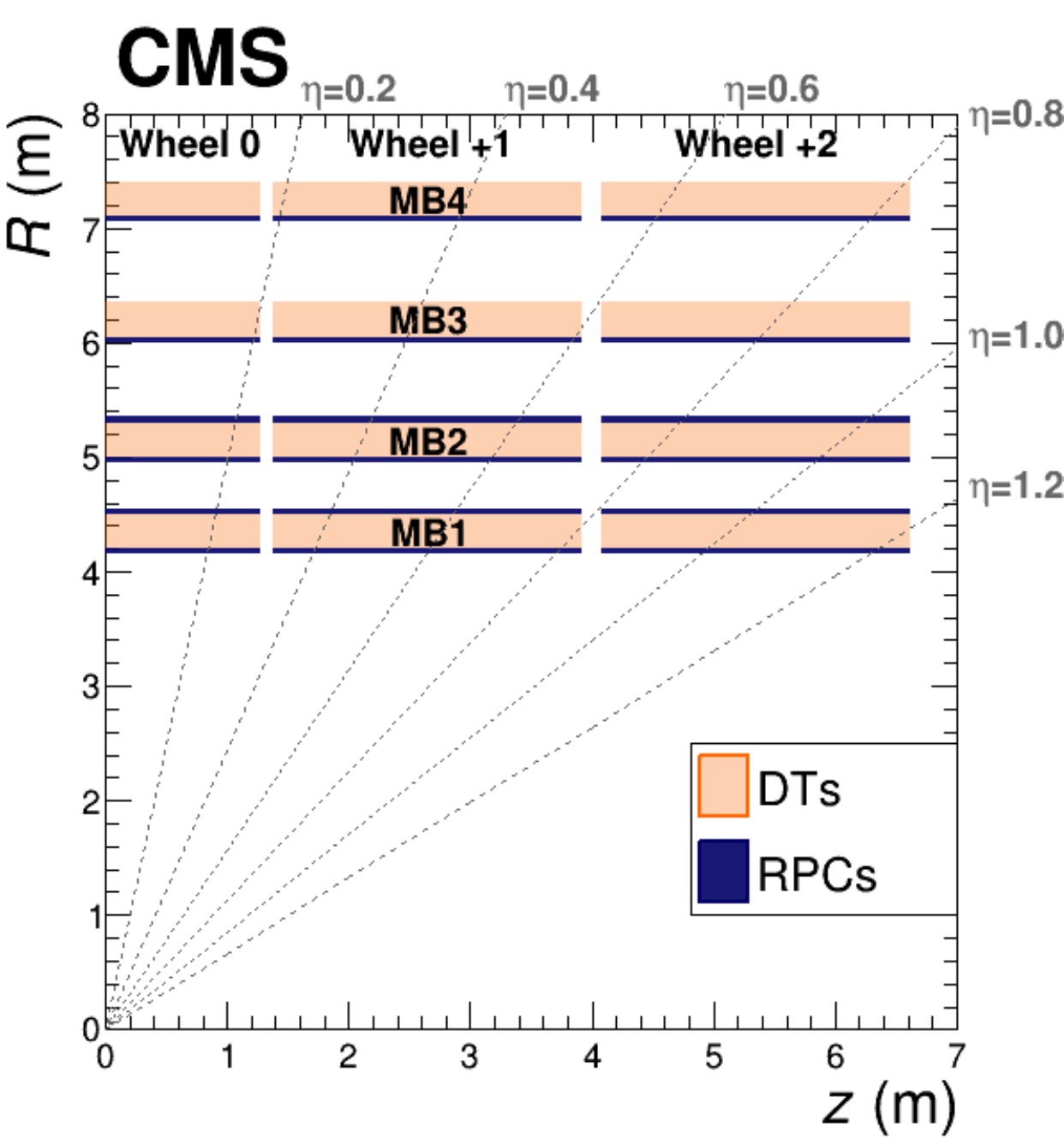}
	\caption{The $R$--$z$ projection of a quarter of the CMS barrel muon system, where $R$ is the radial distance from the beamline.}\label{fig:geometry}
\end{figure}

The L1 muons in the barrel ($\abs{\eta}<0.83$) are reconstructed with the Kalman-filter-based 
barrel muon track finder (kBMTF)~\cite{CMS:2023gfb}, 
which relies on information from the drift tubes (DTs) and resistive-plate chambers (RPCs) of the muon detector in the barrel.
The DT and RPC barrel systems consist of four cylindrical layers wrapped around the solenoid,
called MB1--4 from the innermost to the outermost,
each split into twelve wedges in $\phi$ and five wheels along the beam direction. The distance between MB1 and MB4 is about 3\unit{m}. 
The geometry of the muon detector in the barrel is illustrated in Fig.~\ref{fig:geometry}. The MB1 and MB2 layers both include 
two chambers of RPCs, whereas MB3 and MB4 only have one each.

The TwinMux system~\cite{CMS:2017woh} merges segments in the DTs with RPC hits from the
same layers into objects called stubs, which combine the better
spatial resolution of the DTs with the more precise timing of the RPCs.  The TwinMux processor sends the stubs to the kBMTF,
which builds muon track candidates, assigns a quality to
each, and measures the charge sign and \pt from the bending in the fringe field
of the magnet yoke.
Track quality is related to the estimated \pt resolution, and tracks with more stubs,
particularly those in inner layers where the magnetic
field is stronger, are assigned higher quality. Throughout the rest of the Letter, tracks refer to objects built 
from muon detector information only. 

\section{The L1 data scouting system}

The L1DS system is a novel data collection scheme that acquires
data directly from the L1 trigger, at the full BX rate of 40\unit{MHz}, corresponding to 25\unit{ns} between BXs~\cite{CMS:2023gfb}.
It does not affect the L1 trigger accept decision. 
A subset of objects reconstructed by the L1 trigger (electrons, muons, hadronically decaying $\Pgt$ leptons, jets, energy sums, and muon stubs) 
are stored for events collected by the L1DS, 
corresponding to an average effective record size per BX of about four orders of magnitude 
smaller than that in regular CMS data taking. 
The L1DS system processes and stores data in orbits, each made of 3564 consecutive BXs.  
Unlike traditional data sets where the L1 trigger decision leads to reading out the detector information of a given BX, 
the L1DS system collects
data from all BXs without any preselection. Because of storage limitations, the data are collected with a prescale factor $N$ 
at the orbit level, corresponding to one orbit 
stored every $N$ orbits. Different selection streams with 
requirements on L1 objects also exist, without any prescale factor. 

While the L1DS system will reach its full capacity during the High-Luminosity LHC, a demonstrator started taking data in 2024. 
Since August 2024, stubs from the 
individual layers of the barrel muon detector have been included in the L1DS output for each BX. 
Offline, the stubs from different BXs can be fitted as muon-like tracks, as detailed in Section \ref{sec:reco}, effectively 
enabling the reconstruction of unit-charge particles 
with $0.15\lesssim\beta\lesssim0.80$. 
Below 0.15, the particles mostly get stopped before crossing the muon detector, whereas above 0.80, 
the particle trajectory is reconstructed within a same BX. 
These data were collected with a prescale factor of $N=15$, corresponding to an effective integrated luminosity of 3.7\fbinv, 
from a delivered one of about 55\fbinv.  

\section{Signal and background simulation}\label{sec:samples}

\begin{figure}[htbp]
  \centering
        \includegraphics[width=0.49\textwidth]{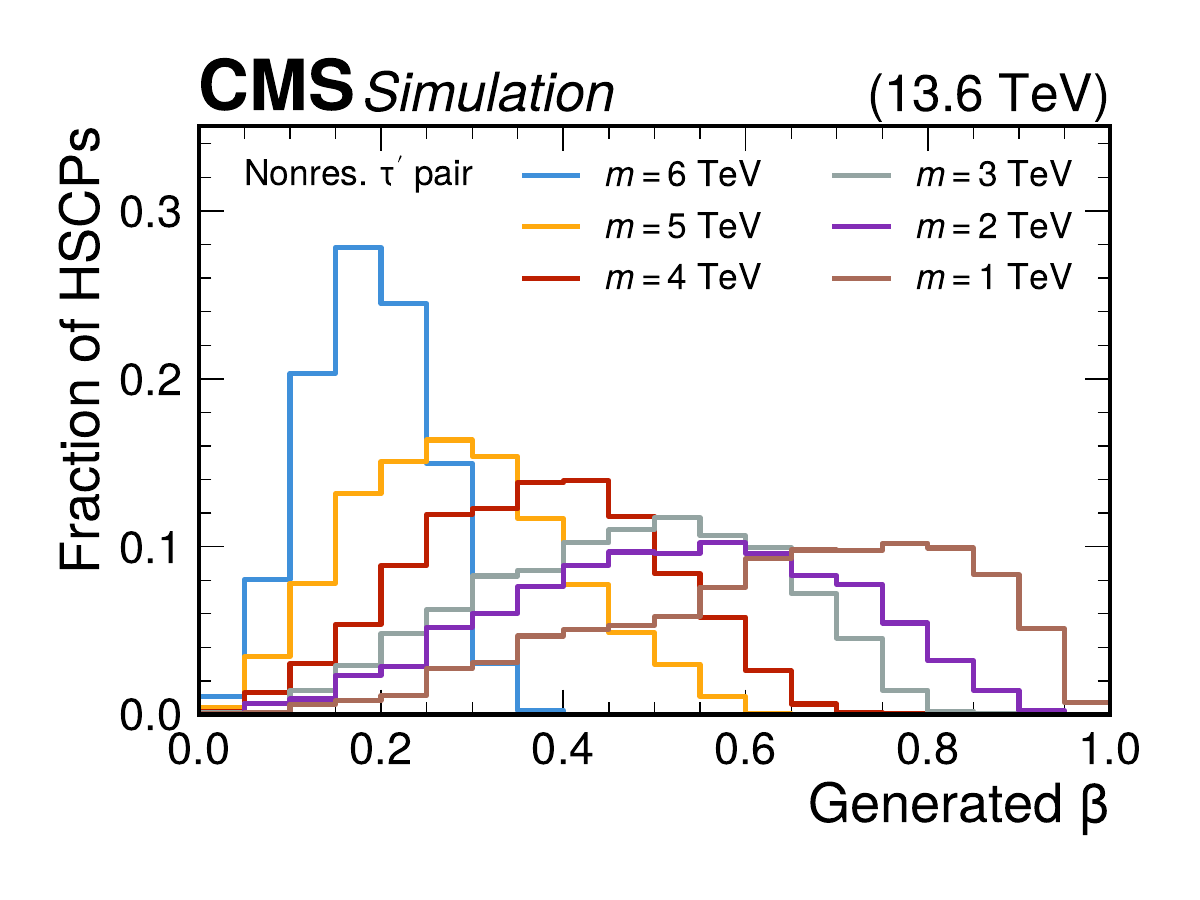}
	\caption{Generated $\beta$ distributions of the fourth-generation $\PGtpr$ leptons from nonresonant DY production, for various $\PGtpr$ masses $m$. The histograms are normalized to unity. 
	}\label{fig:sim}
\end{figure}

Two lepton-like and one hadron-like signal models are used to interpret the results. 
The first lepton-like model consists of modified
Drell--Yan (DY) production of long-lived massive fourth-generation fermions ($\PGtpr$) with spin 1/2 and unit charge, 
as considered in Ref.~\cite{CMS:2024nhn}. As they are neutral under both ${\cal SU}(3)_{\rm C}$ and ${\cal SU}(2)_{\rm L}$ groups,
they couple only to the photon and the $\PZ$ boson. 
Simulated event samples for $\PGtpr$ fermions are produced with masses $m_{\PGtpr}$ ranging from 1.0 to 6.5\TeV. Their generated $\beta$ 
distributions are shown in Fig.~\ref{fig:sim} for different $m_{\PGtpr}$. 
The second lepton-like model consists of resonant production of two $\PGtpr$ leptons from the decay of a heavy $\PZpr$ boson,
using a minimal simplified
model. It is inspired by Ref.~\cite{Giudice:2022bpq} but considers only unit-charge HSCPs. 
Masses of the $\PGtpr$ lepton between 1 and 6\TeV are probed, with the $\PZpr$ boson masses $m_{\PZpr}$ in the range between twice and three times $m_{\PGtpr}$ to stay in the limit of low
$\beta$, with a maximum $m_{\PZpr}$ of 12\TeV. Signal samples with the same ratio between $m_{\PZpr}$ and $m_{\PGtpr}$  
have similar $\beta$ distributions. 

The hadron-like model consists of pair production of long-lived gluinos $\PSg$, which is a common signature of
mini-split supersymmetric models where the $\PSg$ decay is highly suppressed by the large scalar quark 
mass~\cite{Farrar:1978xj,Giudice:2004tc,Arkani-Hamed:2004ymt}. In this analysis, a quasi-stable $\PSg$ is assumed and no $\PSg$
decay is considered. The fraction, $f$, of produced $\PSg$ hadronizing into a $\PSg$-gluon
state ($R$-gluinoball) is a parameter of the hadronization model and corresponds to the fraction of
$R$-hadrons that are neutral at production. 
We choose $f=1$, such that these $R$-hadrons would escape analyses looking for anomalous ionization losses 
in the tracker, while they could leave a
signature in the muon chambers if they become charged when crossing the detector through their interactions with matter, 
simulated using the model from Refs. \cite{Kraan:2004tz,Mackeprang:2006gx}. 
About one third of these $R$-hadrons are charged when they cross the muon detector. Their generated $\beta$
distributions are similar to those shown in Fig.~\ref{fig:sim} for the 
fourth-generation $\PGtpr$ leptons produced in nonresonant DY production.

The leading-order (LO) 
\PYTHIA generator v8.240~\cite{Sjostrand:2014zea} with the tune CP5~\cite{CMS:2019csb} and the parton distribution function set
NNPDF3.1~\cite{Ball:2011uy,Ball:2013hta,Ball:2017nwa} at next-to-next-to-LO 
is used to simulate these three signal models. Backgrounds in the analysis are determined from data, but a DY simulation with 
dimuon decays (DY($\Pgm\Pgm$)) is used to calibrate the reconstructed objects. The \MGvATNLO~2.6.5 event generator is used to generate 
DY($\Pgm\Pgm$) events 
at next-to-LO with the FxFx jet matching and merging~\cite{Frederix:2012ps}.
The response of the CMS detector is simulated with the GEANT4 package~\cite{Agostinelli:2002hh}. The simulated
events are reconstructed using the same algorithms as used for data. In all samples,
simulated minimum bias events are overlaid with the primary collision to produce the effect of
additional $\Pp\Pp$ interactions in the same or adjacent BXs, referred to as ``pileup"~\cite{Lange_2015}.

\section{Muon track reconstruction}\label{sec:reco}

In this analysis, HSCPs are reconstructed as muons from the stubs stored in the L1DS data set, using a modified version of the kBMTF. 
While standard L1 muons are built online from stubs in the same BX, the modified kBMTF algorithm is run 
offline using stubs that are separated in time by at most eight BXs. No
constraint on the time ordering of stubs is applied at this stage to preserve unphysical orderings for background 
estimation purposes, as detailed in Section~\ref{sec:bkg}. The fit is performed independently of the 
BX information of individual stubs. 
The average \pt resolution of the muon tracks 
is measured to be 18\% in a simulated DY($\Pgm\Pgm$) sample. 
While the \pt resolution is independent of $\beta$, it gets worse at significantly higher \pt values. 
For $\pt\gtrsim650\GeV$, the tracks in the muon detector are basically straight and different \pt values 
cannot be distinguished anymore. 

\begin{figure}[tb!]
  \centering
  \includegraphics[width=0.45\textwidth]{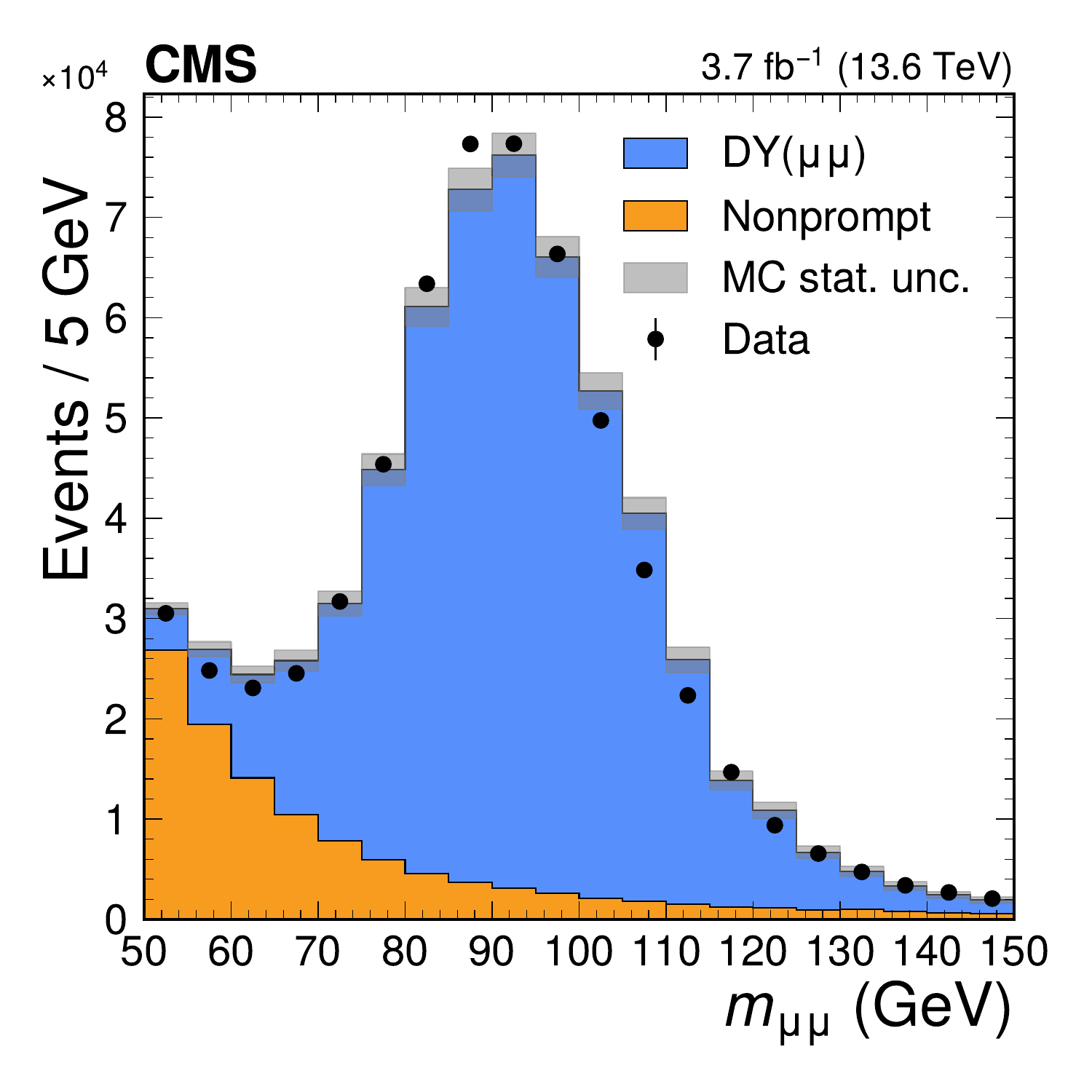}
  \includegraphics[width=0.45\textwidth]{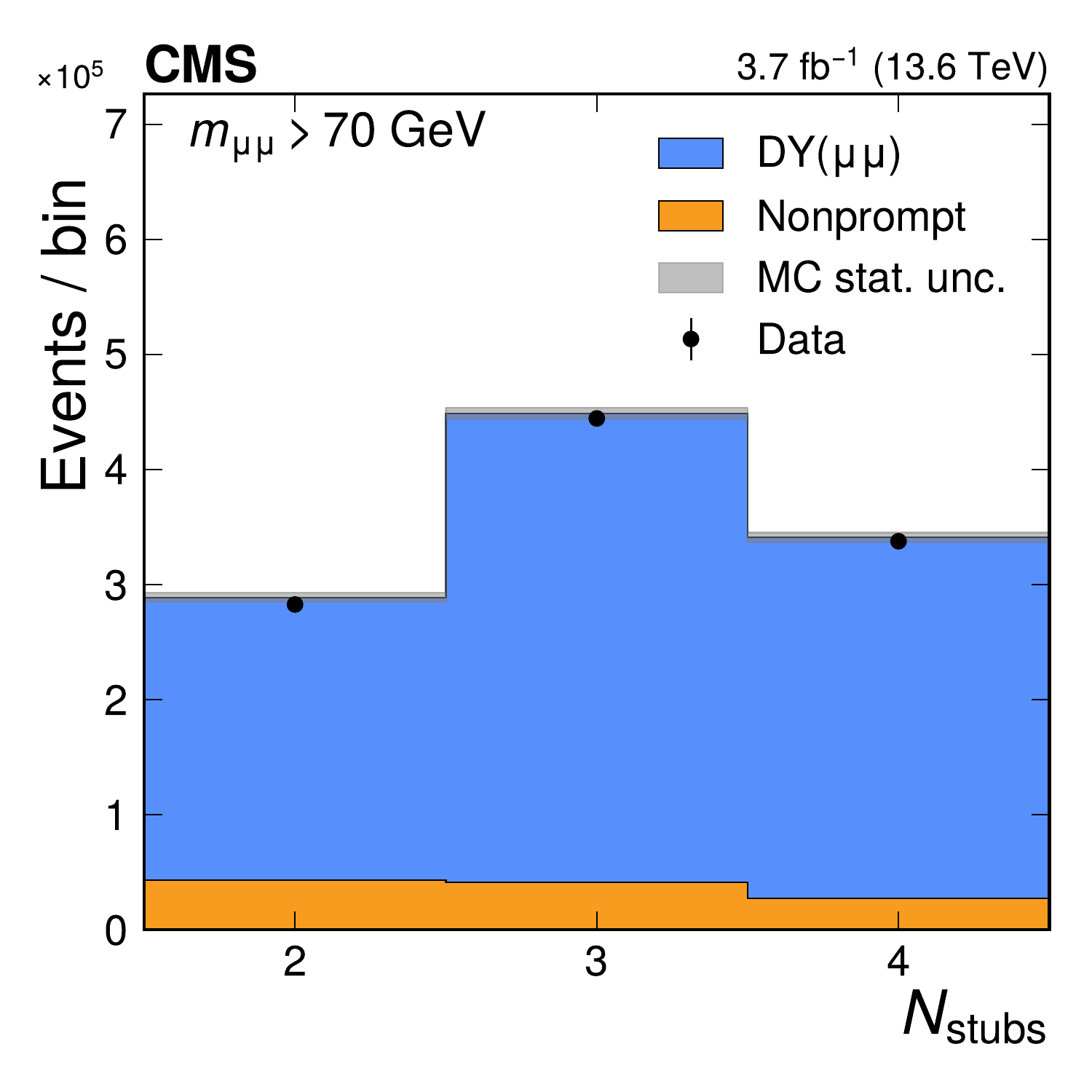}
	\caption{Distributions of the dimuon invariant mass (\cmsLeft) and per-track stub multiplicity (\cmsRight) for events with two opposite-sign modified kBMTF tracks reconstructed from individual stubs in the same or different BXs. The invariant mass is required to be greater than 70\GeV in the stub multiplicity distributions to increase the purity. The muons are selected with $\pt>15\GeV$ and $\abs{\eta}<0.83$. The nonprompt-background contribution is estimated by rescaling the data with two same-sign modified kBMTF tracks. The statistical uncertainty is indicated in the gray shaded area. The agreement between simulation and data is good by construction since the efficiency, energy scale, and energy smearing corrections derived from the same events have been applied.
	}\label{fig:calibration}
\end{figure}

The modified kBMTF reconstruction is validated in
DY($\Pgm\Pgm$) events, which are also used to calibrate the \pt estimation and derive efficiency corrections used for 
simulated samples. The events are selected by requiring a pair of opposite-sign kBMTF tracks with $\pt>15\GeV$ and $\abs{\eta}<0.83$. 
The pair must have an invariant mass $m_{\Pgm\Pgm}$ larger than 50\GeV. 
The nonprompt background, originating mostly from processes like $\PW+\textrm{jets}$ or QCD multijet events, with jets misidentified as 
leptons, is estimated by rescaling the data with same-sign kBMTF tracks to have perfect agreement between data and predictions in 
the opposite-sign events with $50<m_{\Pgm\Pgm}<55\GeV$. 
While the tracks can be built from stubs in different BXs, most DY($\Pgm\Pgm$) tracks are reconstructed within the same BX. 
Five \pt-independent parameters are determined for the simulation: a multiplicative factor of the predicted \pt (compatible with 1.0), the width of a Gaussian smearing of the predicted \pt (about 10\%), and an efficiency correction per stub multiplicity (up to 11\% for 4-stub tracks).    
Figure \ref{fig:calibration} shows the dimuon invariant mass and the stub multiplicity per muon track after 
the corrections have been applied, where good agreement between predictions and data is obtained by construction. 
In these events, about 25 (45, 30)\% of the reconstructed muons have 2 (3, 4) stubs.

\begin{figure}[tb!]
  \centering
  \includegraphics[width=0.49\textwidth]{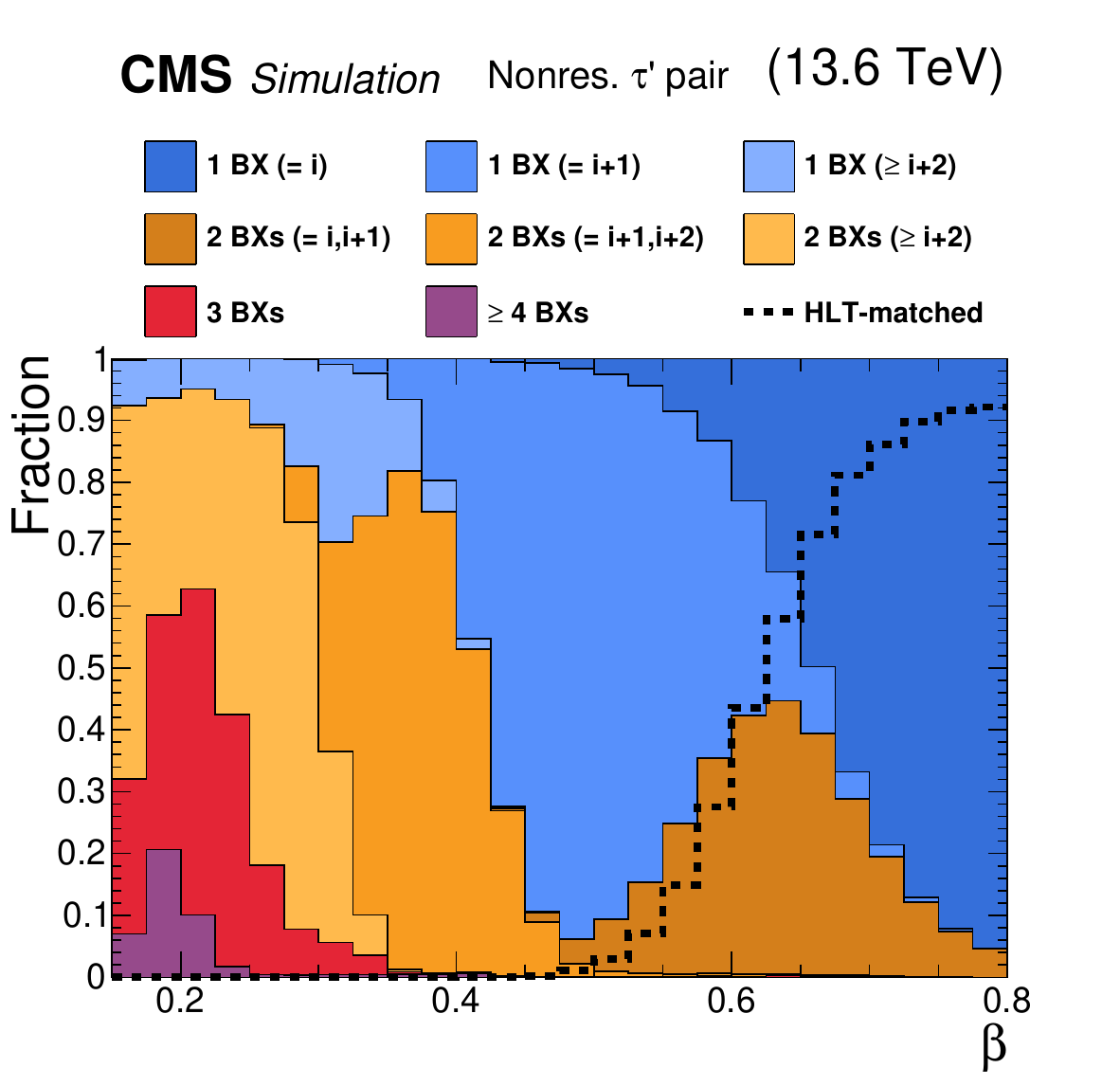}
        \caption{
                Fractions of the BX signatures of tracks reconstructed with the modified kBMTF algorithm as functions of the generated particle $\beta$. This figure corresponds to the nonresonant fourth-generation lepton signal model with a mixture of HSCP masses between 1 and 6\TeV. The filled histograms are stacked to sum to unity. The blue histograms correspond to tracks reconstructed in a single BX. Different shades represent different BXs with respect to the collision BX ($\mathrm{BX} = i$), as indicated in parentheses in the legend. The orange histograms correspond to tracks reconstructed over 2 BXs, with different shades corresponding to different BXs with respect to the collision BX. Different shades of the same color are indistinguishable experimentally since the production BX cannot be determined. The red and purple histograms represent tracks reconstructed over 3 or at least 4 BXs, respectively. The dashed line indicates the efficiency for the reconstructed tracks to satisfy the single-muon HLT selection. For $\beta<0.15$, the efficiency goes to zero because the particles get stopped in the detector. For $\beta>0.8$, the particles are increasingly reconstructed in the same BX, in which case the HLT selection is fully efficient.
	}\label{fig:fraction}
\end{figure}

Depending on their $\beta$, HSCPs are reconstructed within different ranges of BXs. Figure \ref{fig:fraction} 
shows the fraction of the tracks reconstructed in different BX signatures for unit-charge HSCPs. For $\beta>0.7$, particles 
mostly cross the muon detector within the same BX as the hard scattering (dark blue). 
The same experimental signature of stubs within the same 
BX appears for most HSCPs with $\beta$ around 0.5, but in that case the particle takes one BX to reach the first layer of the muon 
detector then crosses all the layers within the BX following that of the collision (lighter blue). Up to 30\% of 
HSCPs with $\beta$ around 0.3 also have the same experimental signature, but with the muon detector layers crossed at least 2 BXs after 
the hard scattering (lightest blue). About half of the 
HSCPs with $\beta=0.6$ cross the muon detector during the collision BX and the following one (dark orange). 
HSCPs with $\beta=0.4$ mostly take one BX to reach the first layer of the muon detector, and reach the last muon detector layer one BX later 
than the first layer (lighter orange), whereas a significant fraction of HSCPs with $\beta\lesssim 0.3$ need at least 2 BXs to reach the first 
layer, and cross the last layer one BX later (lightest orange). 
Different shades of a same color, corresponding to different BX intervals with respect to the collision BX, 
are indistinguishable experimentally since the production BX cannot be determined.
For $\beta\lesssim 0.3$, a significant 
fraction of the tracks are reconstructed across 3 or 4 BXs (red or purple histograms, respectively), whereas the efficiency goes to zero 
for $\beta\lesssim 0.15$ because the particles get stopped in the detector.
For $\beta\lesssim0.5$, the efficiency for the reconstructed tracks to satisfy the single-muon high-level trigger (HLT) selection is zero, 
showing that such slow particles cannot be directly triggered on.

\section{Event selection, categories, and statistical analysis}

Events containing at least one modified kBMTF track with 3 or 4 stubs,
reconstructed across a minimum (maximum) of 2 (4) BXs, are selected. Tracks with 2 stubs are
not considered because of the relatively larger background.
Particles slow enough to cross the muon chambers in more than 4 BXs would mostly be stopped in the detector 
because of their large ionization losses.
In the signal region (SR),
hits must occur from the inner layers to the outer ones, \ie, a stub in an outer layer must be in the same BX or a later BX with respect to
the stub in the closest inner layer. The muon tracks must have $\abs{\eta}<0.83$ and $\pt>15\GeV$. 
If several such tracks are present, the slowest track (\ie, spanning the most BXs) is selected.
If speeds are equal, the track with the largest stub multiplicity is chosen; 
if both the speed and stub multiplicity are equal, the highest \pt track is selected.
The selected track is required to have a transverse distance of closest approach to the beamline center, $d_{xy}$, less than 32\cm, 
and to pass high-quality criteria calculated by the kBMTF algorithm. 
The coarse selection on $d_{xy}$ is related to the low precision available from the 
L1 information and algorithms. 
Colliding bunches are determined per orbit on the basis of the LHC injection scheme. 
Events are discarded if the BX of the earliest stub and the five BXs preceding it are all noncolliding, since HSCPs cannot
be produced in noncolliding bunches.
For the simulated samples of the nonresonant DY production of $\PGtpr$ leptons, the product of the acceptance and efficiency of 
these selection criteria is 19\% for $m_{\PGtpr} = 1\TeV$ and as high as 47\% for $m_{\PGtpr} = 5.5\TeV$. 

Fourteen categories are built to target signals with different $\beta$, characterized by different crossing times of the detector. The categories
are also based on the stub multiplicity since the signal-to-background ratio improves with the number of stubs.
The categories are defined in Table \ref{tab:categ}. The most sensitive category, BX123, is illustrated in 
Fig.~\ref{fig:BX123}, whereas the illustrations of all categories can be found in \cmsApp\ref{app:schematic}.

\begin{figure*}[htbp]
  \centering
        \includegraphics[width=0.75\textwidth]{Figure_005.pdf}
        \caption{Schematic definition of the BX123 category. 
}\label{fig:BX123}
\end{figure*}

\begin{table*}[htb!]
        \centering
	\topcaption{Categories based on muon detector layers with stubs and speed expressed as the number of crossed layers divided by the number of BXs spent to cross them. The BX1234 category includes 4-stub tracks for which the 4 stubs are in 4 subsequent BXs. For 3-stub (4-stub) tracks, the BX123 categories correspond to 3 (4) stubs in 3 different BXs across a range of 3 BXs, whereas the BX124/134 categories correspond to 3 (4) stubs in 3 different BXs across a range of 4 BXs. The 4-stub category BX1112 (1122, 1222) corresponds to 3 (2, 1) stubs in the first BX, and 1 (2, 3) stubs in the next BX. Similarly, the 3-stub categories BX112 (122) correspond to 2 (1) stubs in the first BX and 1 (2) stub in the next BX. The 3-stub categories are further separated into fast and slow subcategories: those denoted ``fast" feature a track with the first stub in MB1 and the last one in MB4, corresponding to a longer distance crossed in the same amount of time as their ``slow" counterparts, which do have exactly one stub within the innermost and outermost detector layers. The symbol $\oplus$ represents the exclusive ``or", whereas $\land$ is the logical ``and". }\label{tab:categ}
        \begin{tabular}{lccccc}
		Stub config.         & \multicolumn{5}{c}{Layers / BXs}  \\
		                     & 3/4            & 3/3 or 4/4     & 4/3        & 3/2                 & 4/2 \\
                \hline
                4 stubs              & $\NA$          & BX1234         & BX123      & $\NA$               & BX1112 \\
                                     &                & BX124/134      &            &                     & BX1122 \\
                                     &                &                &            &                     & BX1222 \\ [\cmsTabSkip]
                3 stubs,             & BX124/134 slow & BX123 slow     & $\NA$      & BX112 slow          & $\NA$     \\
                MB1 $\oplus$ MB4           &                &                &            & BX122 slow          &        \\ [\cmsTabSkip]
                3 stubs,             & $\NA$          & BX124/134 fast & BX123 fast & $\NA$               & BX112 fast\\
                MB1 $\land$ MB4          &                &                &            &                     & BX122 fast\\
        \end{tabular}
\end{table*}

The categories with tracks across 2 BXs, namely with 3/2 or 4/2 layers/BXs, 
suffer from a large background from per-stub BX misidentification, which is below 1\% for muon detector layers 
with two working layers of RPCs, and up to 6\% when no layer of RPCs is working~\cite{CMS-DP-2025-008}. 
Additional model-dependent selection criteria based on the pair-production hypothesis can significantly reduce the 
background. In these categories, a second good-quality track with $\abs{\eta}<0.83$ and $\pt>100\GeV$ is required, with at least one stub in the same BX as one of the stubs of the first track.  
This reduces the background by almost two orders of magnitude, while retaining about half of the signal.
The original categories without the additional selection criterion are kept to set model-independent limits on the fiducial cross section of slow particles, without relying on the  
description of the rest of the event.

The results are extracted with a maximum likelihood fit of binned distributions in each category. 
The observable is chosen to be the track \pt because the signal is characterized by harder objects than the background. 
Indeed, particles with low $\beta$ reaching the muon chambers must be heavy and produced with significant momentum. 
To avoid overconstraining the background estimate from low-\pt bins with large event yields, the \pt threshold is increased to 50 (150)\GeV 
in the categories with tracks across 2 BXs with (without) a requirement on the presence of a second track, and to 50\GeV in 
all categories across more than 2 BXs, except the BX1234 category where the \pt threshold is 15\GeV because 
of the lower background expectation.

\section{Background estimation and validation} \label{sec:bkg}

Backgrounds arise mostly from ultra-relativistic particles with BX misidentification in the individual muon detector layers, and 
from combining unrelated ultra-relativistic particles from different BXs at the same place in the detector,
but also from noncollision processes and noise in the detector. 
Contributions from slow particles originating from the experimental cavern are negligible because the time ordering of their stubs is 
not compatible with a particle produced at the beam spot and exiting the CMS detector. 
The track \pt distribution of the backgrounds in each category is 
estimated using tracks with an asynchronous time ordering of the stubs within the muon detector layers. The synchronous
signal-like ordering is when the stub in each layer occurs in the same or a later BX compared to the stubs in the layers inside it. 
The other synchronous ordering, where the inner stubs occur later, is not signal-like and signifies a track that enters the 
detector from the outside.
Other combinations are asynchronous.
Because of BX misidentification, some asynchronous combinations include significant contributions from physical processes entering the detector, such as slow neutrons originating from the experimental cavern, which may have a different \pt distribution
from that of the backgrounds in the SR. 
To remove this contribution, only the tracks where either the innermost stub is in the earliest  
BX or the outermost stub in the latest BX are used to estimate the 
background \pt distribution in the SR. 
Signal contributions to events with asynchronous orderings are below the percent level and therefore are neglected. 
The configurations considered to estimate the background in the 
BX1234 and BX1112 categories are illustrated in \cmsApp\ref{app:schematic}. 
The background normalization is left floating in each category in the binned maximum likelihood fit, 
where it gets constrained by the signal-depleted low-\pt bins.

\begin{figure*}[htbp]
  \centering
  \includegraphics[width=0.9\textwidth]{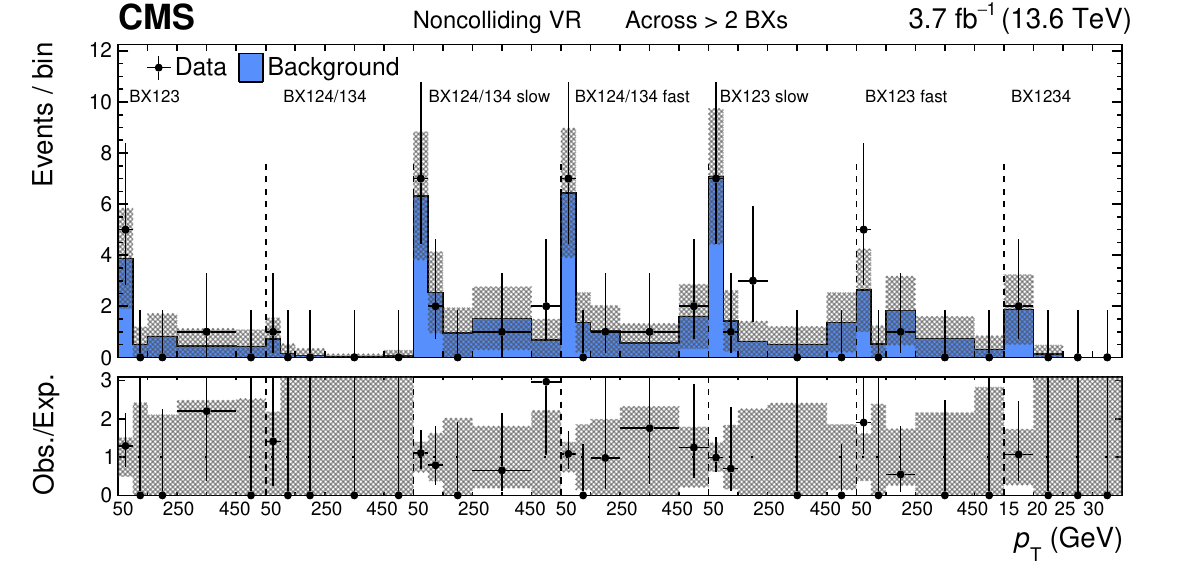}\\
  \includegraphics[width=0.9\textwidth]{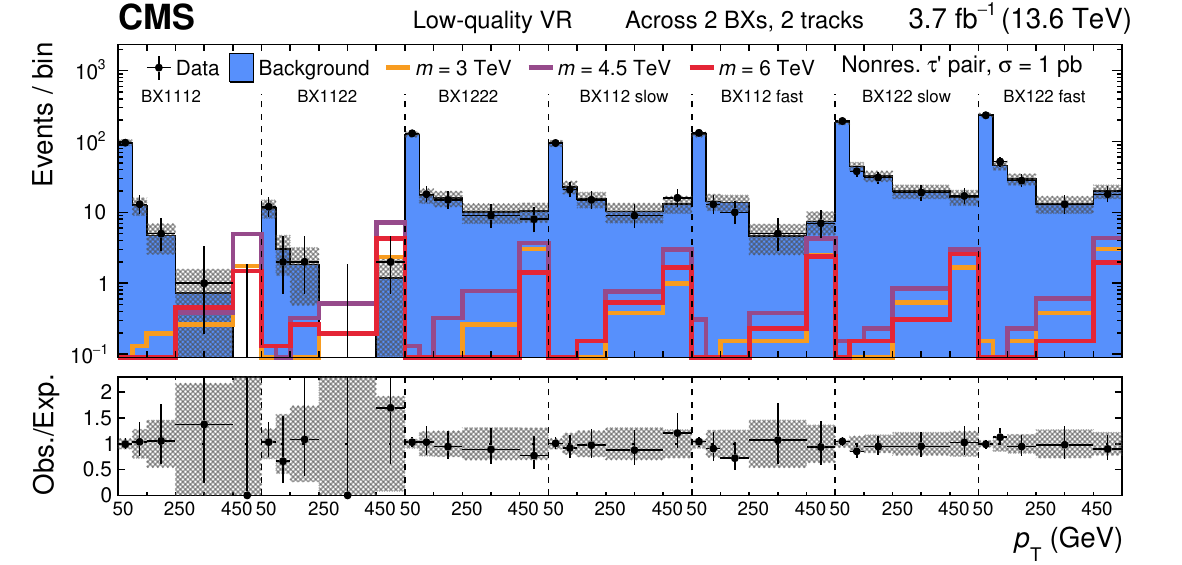}\\
  \includegraphics[width=0.9\textwidth]{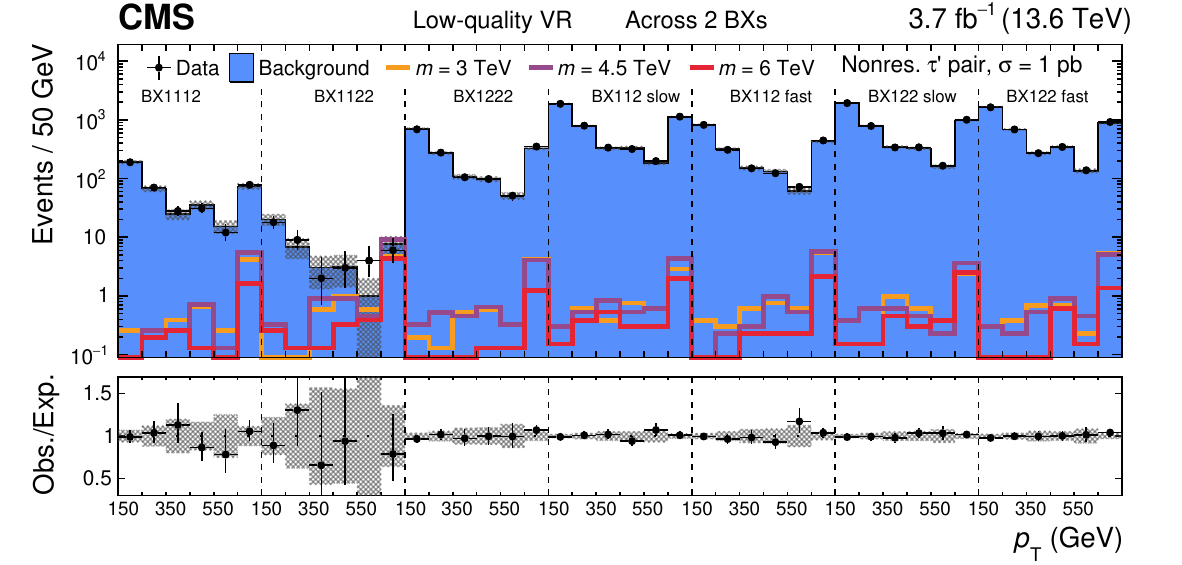}\\
	\caption{Low-track-quality validation region distributions for the categories with tracks across $>$2 ($=$2, $=$2) BXs without (with, without) additional track requirement are shown in the upper (center, lower) part of the figure. The expected background distributions are the result of the maximum likelihood fit. The uncertainty bands account for all sources of background uncertainty, systematic as well as statistical, after the maximum likelihood fit. The signal is shown for a few mass hypotheses of the nonresonant fourth-generation lepton model, using a production cross section of 1\unit{pb}. The lower panels show the observed to expected ratio. The last \pt bins include the overflow.}\label{fig:failingVR}
\end{figure*}

\begin{figure*}[htbp]
  \centering
  \includegraphics[width=0.9\textwidth]{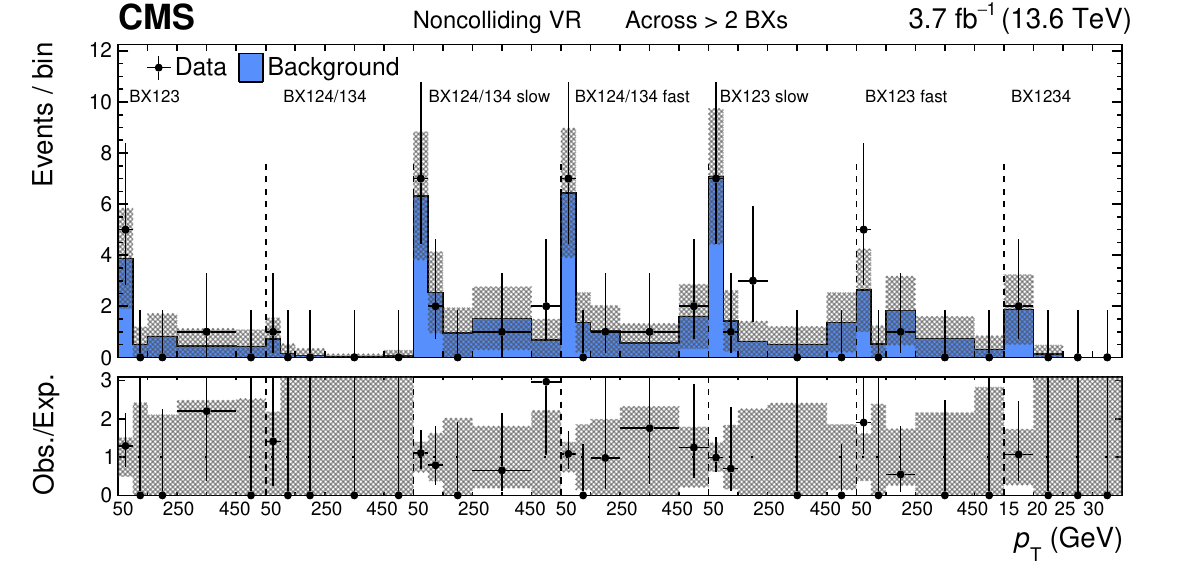}\\
  \includegraphics[width=0.9\textwidth]{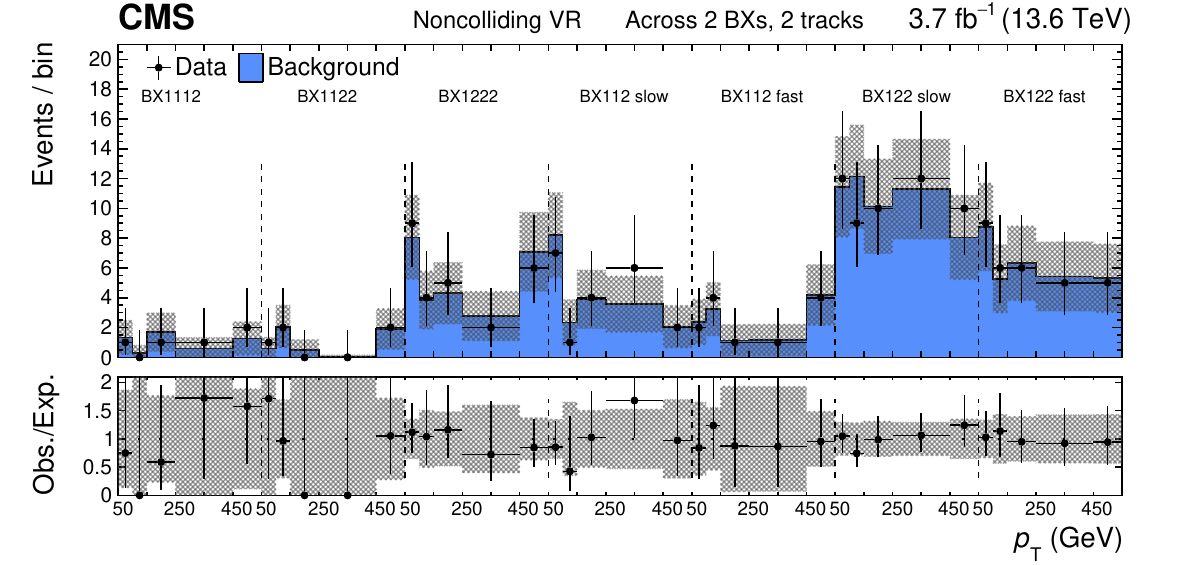}\\
  \includegraphics[width=0.9\textwidth]{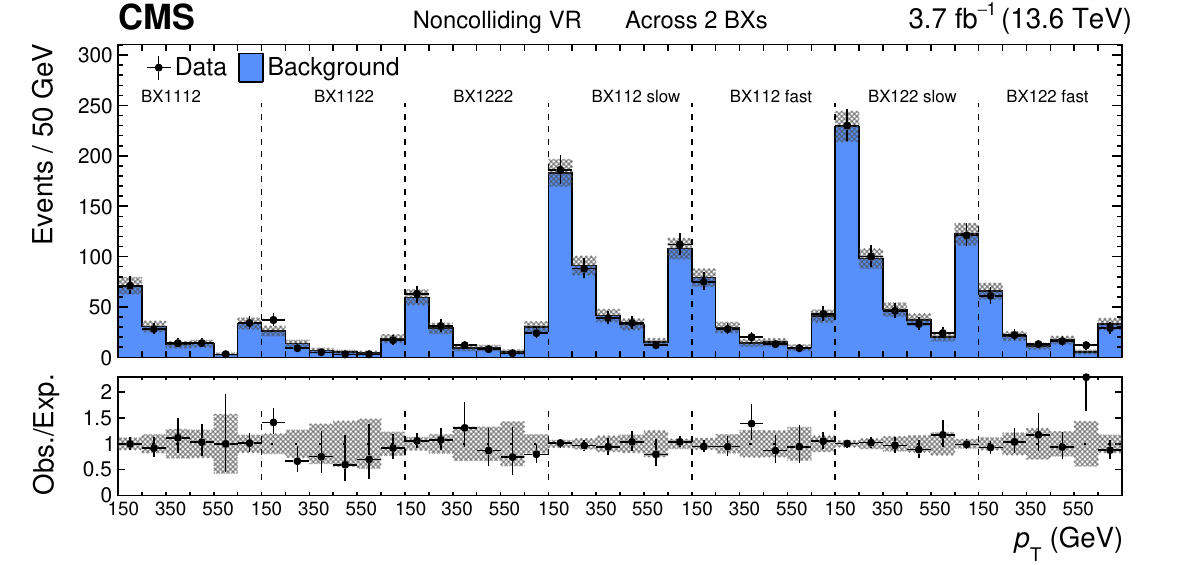}\\
        \caption{Noncolliding validation region distributions for the categories with tracks across $>$2 ($=$2, $=$2) BXs without (with, without) additional track requirement are shown in the upper (center, lower) part of the figure. The expected background distributions are the result of the maximum likelihood fit. The uncertainty bands account for all sources of background uncertainty, systematic as well as statistical, after the maximum likelihood fit. The lower panels show the observed to expected ratio. The last \pt bins include the overflow.}\label{fig:noncollidingVR}
\end{figure*}

The background estimation is verified in two orthogonal validation regions (VRs): (i) events in colliding bunches where 
the track fails the quality criteria, and (ii) events with a high-quality track
where the BX of the earliest stub and the five BXs preceding it are noncolliding. A maximum likelihood fit to the data is performed 
using the systematic uncertainties described in Section \ref{sec:syst}. In both VRs, good agreement between the observed data
and predicted background is observed, as shown in Figs.~\ref{fig:failingVR} and \ref{fig:noncollidingVR}.
The VR with low track quality criteria is significantly more populated than the SR, making it a stringent test of the
background estimation method. The noncolliding VR is significantly less populated than the SR, but provides a useful validation for events
with signal-like high-quality tracks.

\section{Systematic uncertainties}\label{sec:syst}

The dominant systematic uncertainties come from the background estimation. While the background normalization is left floating in the 
fit independently for each category, one systematic uncertainty affecting the \pt distribution is considered in each category. 
These uncertainties are incorporated in the likelihood as nuisance parameters with a Gaussian prior. 
The alternative \pt distributions corresponding to a $\pm1$ standard deviation (s.d.) are built from either a subset of the tracks used 
to estimate the nominal \pt distribution, or from an orthogonal set, as described in the next paragraphs and summarized
in Table \ref{tab:syst}.

In the categories with tracks across more than 2 BXs, as well as in BX1122,
the \pt-dependent systematic uncertainty corresponding to a $+1$ ($-1$) s.d. is determined by considering uniquely the subset of
asynchronous orderings
where the innermost (outermost) stub is in the earliest (latest) BX. 
This results in an uncertainty in the range 5--60\% for tracks with $\pt>450\GeV$. 

This method does not work in
the other categories because these two asynchronous orderings do not both exist or are not distinct.
In BX1112 and BX1222, the alternative \pt distribution
corresponding to a $+1$ ($-1$) s.d. is taken from the subset of asynchronous orderings where
the stub detected in a different BX than the
other three stubs is in MB2 (MB3),
leading to normalization
effects of 5--30\% for events with tracks with $\pt>450\GeV$. 

In the remaining 3-stub categories spanning exactly 2 BXs,
namely BX112 slow, BX112 fast, BX122 slow, and BX122 fast, 
the shape uncertainty is obtained by considering the \pt distribution
of a different set of tracks with asynchronous orderings, failing
the quality criteria, and symmetrizing it with respect to the nominal distribution. It has an effect of 5--75\% for events with tracks with $\pt>450\GeV$. 

The background \pt-dependent uncertainties on average get constrained by the maximum likelihood fit to 50\% of their initial 
magnitudes. Overall, systematic uncertainties have a small impact on the analysis, which is statistically limited by the number of 
events selected in the high-\pt bins of the SR.

\begin{table*}[ht!]
	\topcaption{Definition of asynchronous track orderings used to build the alternative \pt distributions corresponding to a $\pm 1$ s.d. of the background systematic uncertainty.}\label{tab:syst}
        \centering
	\begin{tabular}{lll}
		Categories & $+1$ s.d. & $-1$ s.d. \\ \hline
		BX1122, BX123, BX124/134,       & Asynchronous orderings           & Asynchronous orderings   \\
		BX124/134 slow, BX124/134 fast, & where the innermost stub         & where the outermost stub \\
		BX123 slow, BX123 fast, BX1234  & is in the earliest  BX           & is in the latest BX      \\ [\cmsTabSkip]
		BX1112, BX1222                  & Asynchronous orderings           & Asynchronous orderings   \\
						& where the stub detected in       & where the stub detected in \\
						& a different BX is in MB2         & a different BX is in MB3   \\ [\cmsTabSkip]
		BX112 slow, BX112 fast,         & Asynchronous orderings           & Symmetric distribution     \\
		BX122 slow, BX122 fast          & failing the quality criteria     & with respect to nominal    \\
	\end{tabular}
\end{table*}

The signal normalization uncertainty, coming from the selection efficiency, as well as the integrated luminosity, 
is determined from the efficiency correction derived from DY($\Pgm\Pgm$) events,
which includes effects related to the reconstruction and identification of muon-like particles, as well as overall normalization effects,
such as the DY($\Pgm\Pgm$) cross section and the integrated luminosity. An additional uncertainty comes from the limited size of the 
DY($\Pgm\Pgm$) simulation
used to derive the correction. This results in an uncertainty of about 3 (2)\% in categories with tracks with 3 (4) stubs,
partially correlated between stub
multiplicity categories.

\section{Results}

\begin{figure*}[htbp]
  \centering
  \includegraphics[width=0.8\textwidth]{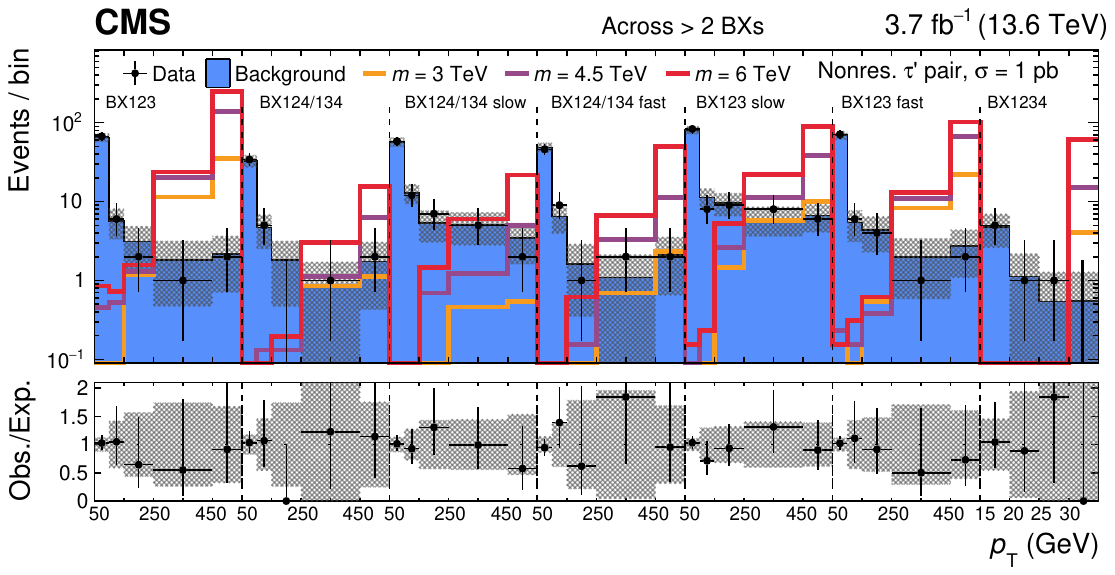}\\
  \includegraphics[width=0.8\textwidth]{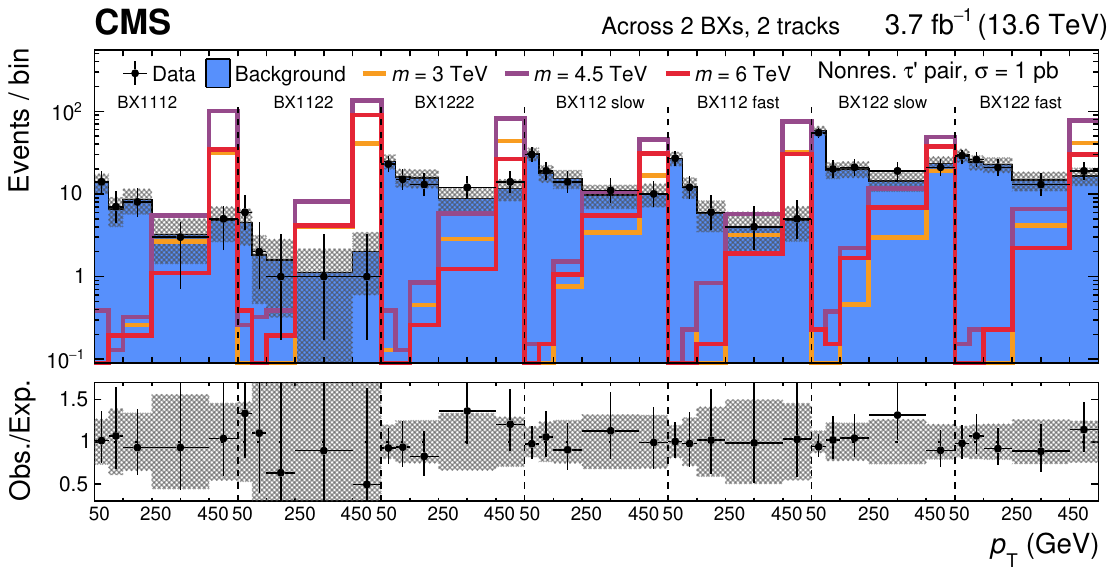}\\
  \includegraphics[width=0.8\textwidth]{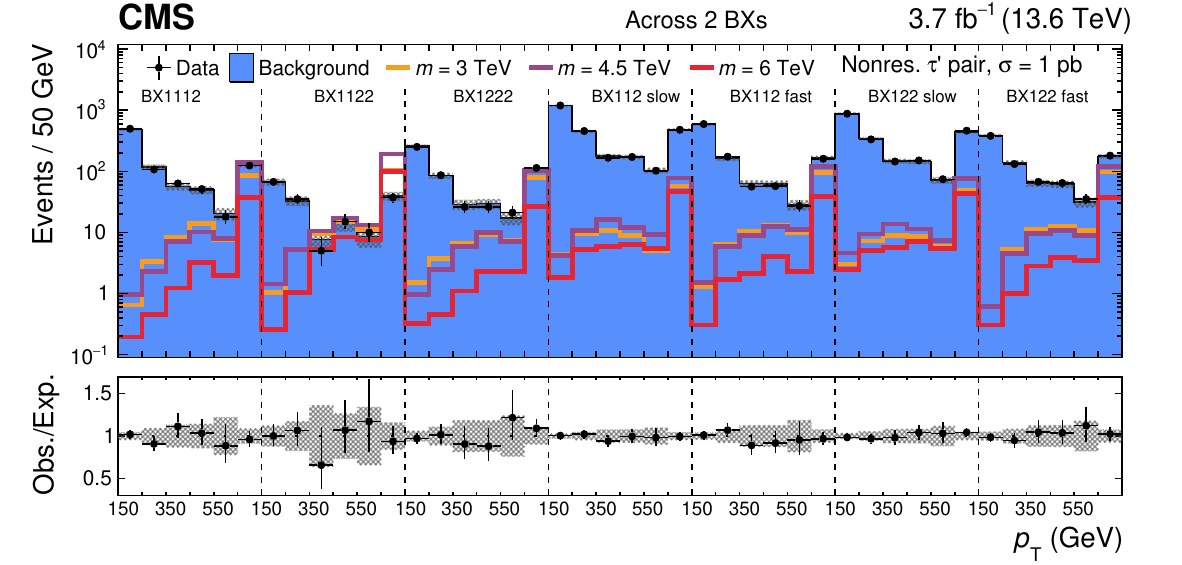}\\
	\caption{The \pt distributions in the signal region for the categories with tracks across $>$2 ($=$2, $=$2) BXs without (with, without) additional track requirement are shown in the upper (center, lower) part of the figure. The expected background distributions are the result of the maximum likelihood fit. The uncertainty bands account for all sources of background uncertainty, systematic as well as statistical, after the maximum likelihood fit. The lower panels show the observed to expected ratio. The signal is shown for a few mass hypotheses of the nonresonant fourth-generation lepton model, using a production cross section of 1\unit{pb}. The last \pt bins include the overflow.}\label{fig:SR}
\end{figure*}

The results are extracted with a binned maximum likelihood fit to the observed \pt distributions in the
different SRs. The systematic uncertainties described in Section \ref{sec:syst} are 
considered as nuisance parameters in the statistical procedure. 
The results have been determined using the CMS statistical analysis tool \textsc{Combine}~\cite{CMS:2024onh}. The SR distributions are shown in Fig.~\ref{fig:SR}. No significant excess of data is observed above the standard model expectation. 

\begin{figure*}[ht!]
  \centering
	\includegraphics[width=0.45\textwidth]{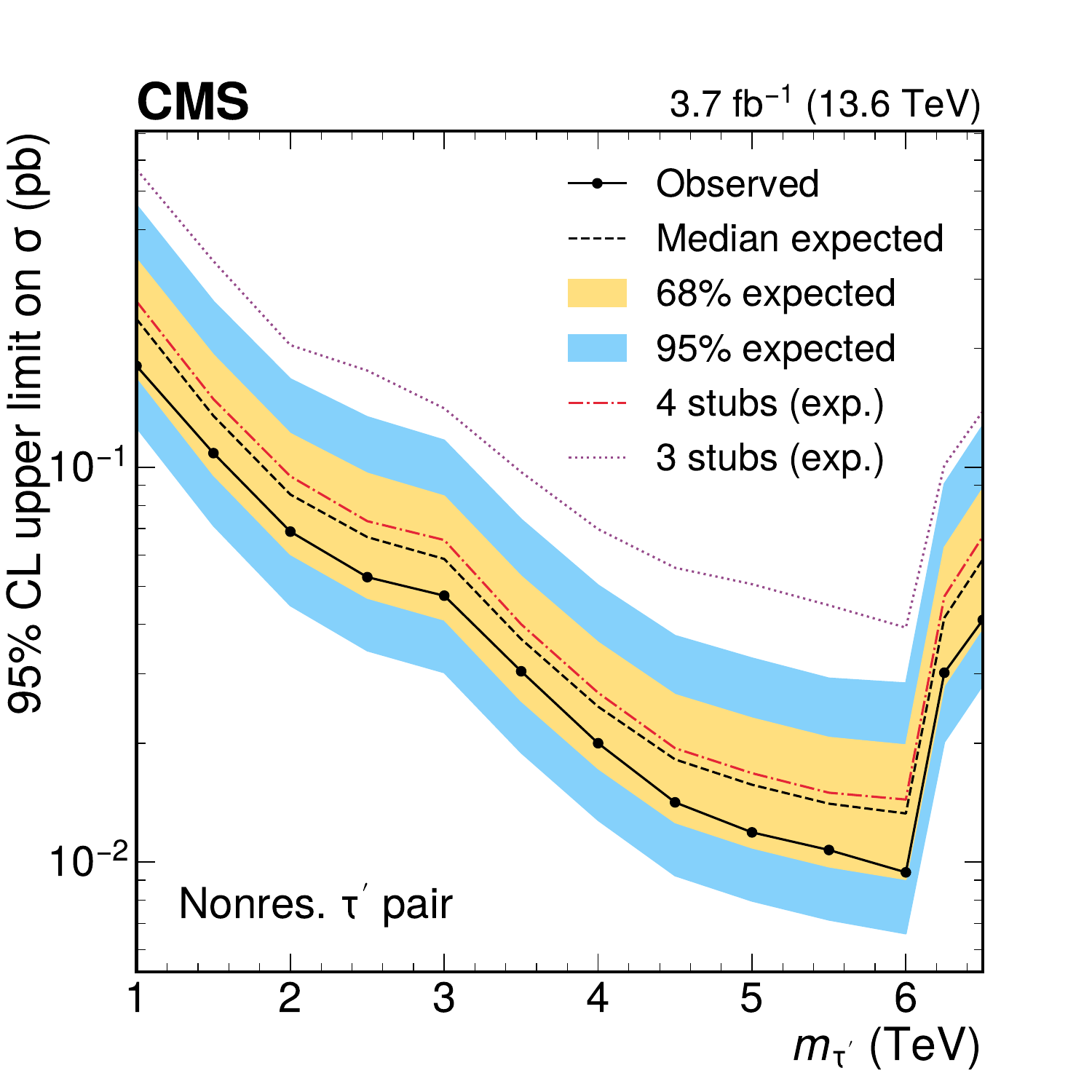}
  \includegraphics[width=0.45\textwidth]{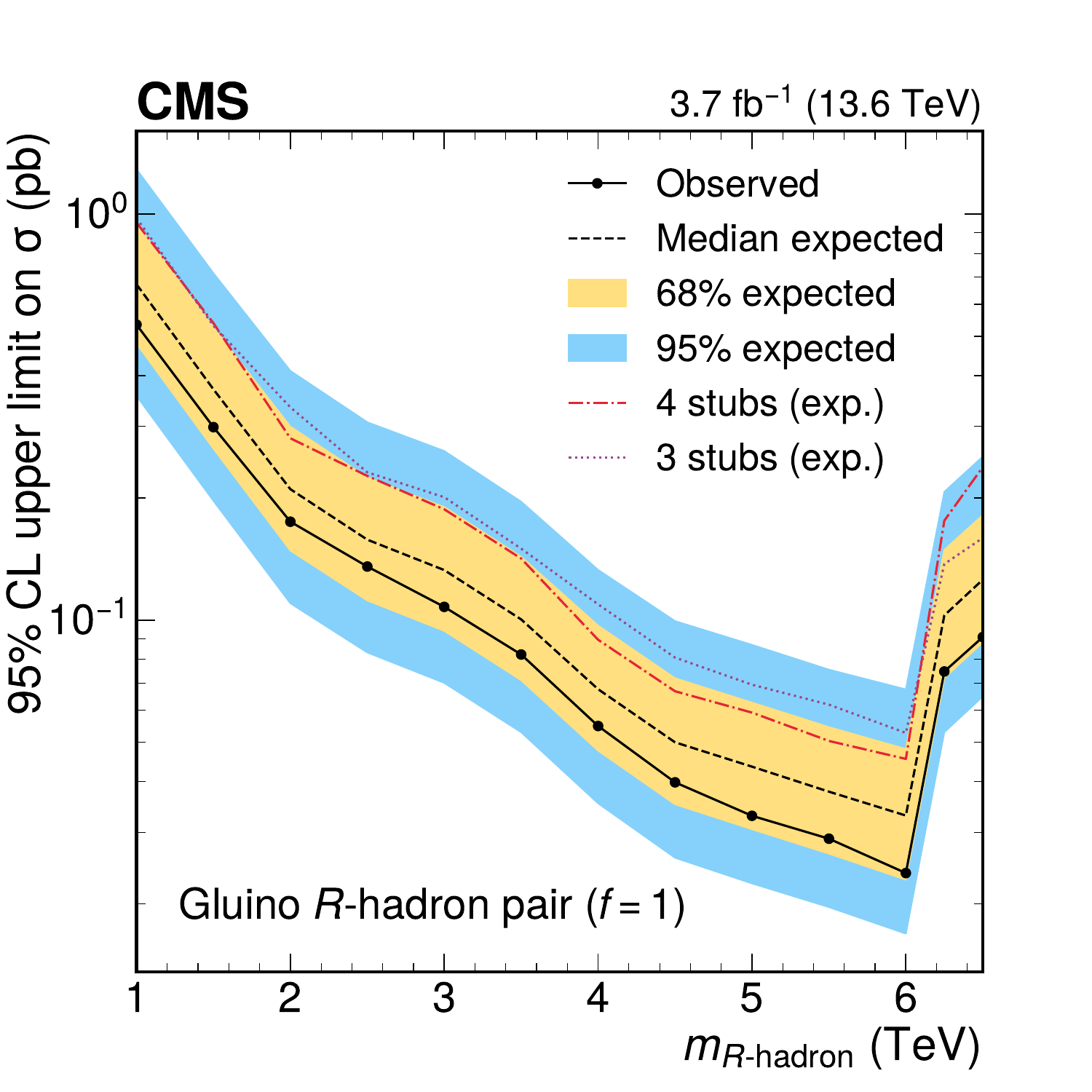}
	\caption{Observed (solid line with markers) and expected (dashed black line) upper limits at 95\% \CL on the production cross section of heavy fourth-generation leptons through nonresonant DY production (left) and of a gluino $R$-hadron pair with $f=1$ (right). The inner (yellow) and the outer (blue) bands indicate the regions containing 68 and 95\%, respectively, of the distribution of limits expected under the background-only hypothesis. The expected limits for the combinations of 3-stub (dashed purple line) and 4-stub (dashed-red line) categories are also shown.}\label{fig:limits}
\end{figure*}

\begin{figure}[htbp]
  \centering
  \includegraphics[width=0.45\textwidth]{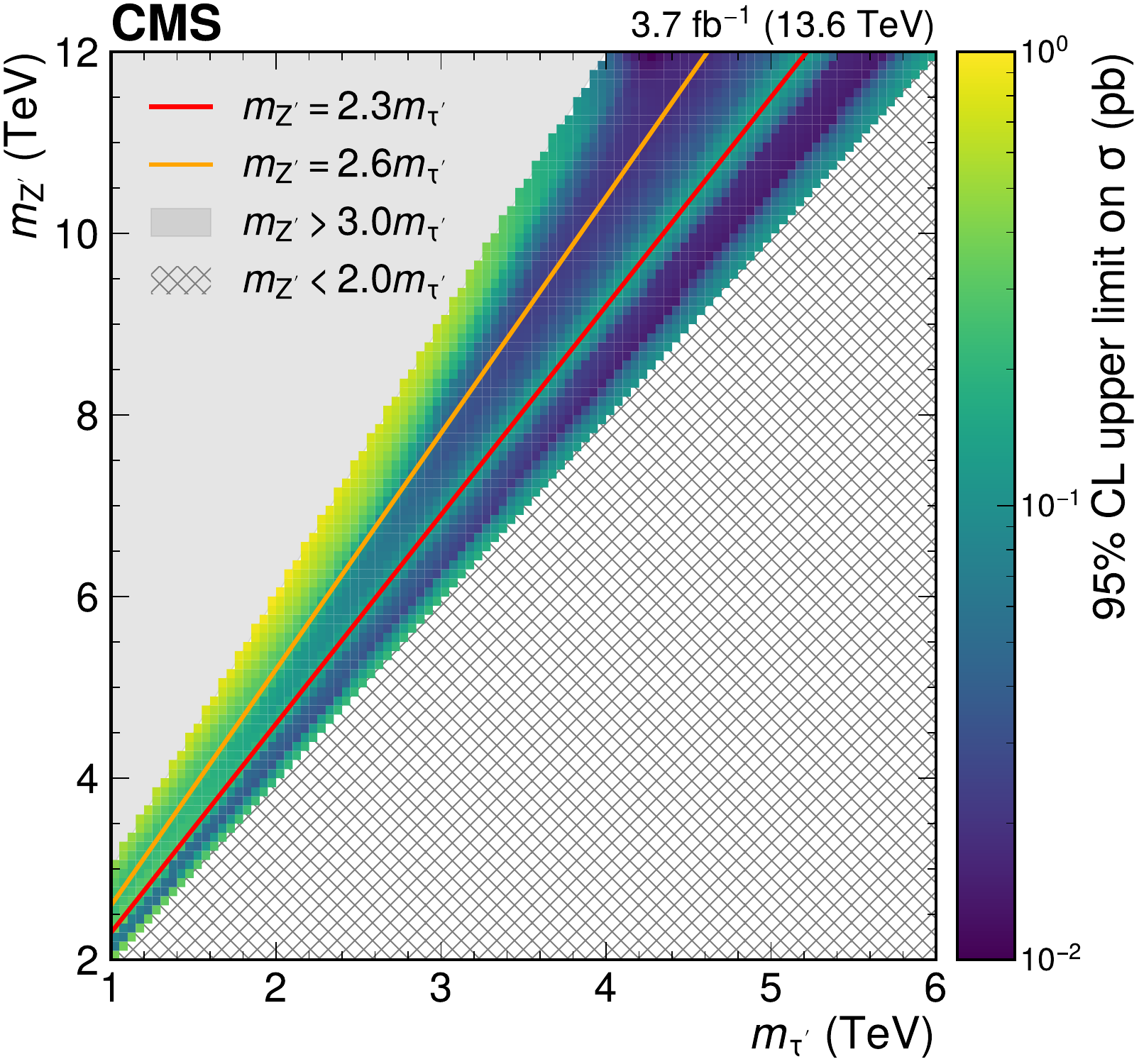}\\
	\caption{Observed upper limits at 95\% \CL on the production cross section of heavy fourth-generation leptons through $\PZpr$ boson decays. Lines indicate different values of the ratio of $\PZpr$ to $\PGtpr$ masses, for which $\beta$ values are typically similar. Signal hypotheses with $2m_{\PGtpr}<m_{\PZpr}<3m_{\PGtpr}$ are probed. For $m_{\PZpr}>3m_{\PGtpr}$, the HSCPs typically have high $\beta$ and are reconstructed in the same BX. The lowest limits are obtained for $m_{\PZpr}/m_{\PGtpr}\approx2.15$ (2.60), for which HSCPs are dominantly produced with $\beta\approx 0.2$ (0.6). 
	The higher limits for $m_{\PZpr}/m_{\PGtpr}\approx2.30$ are related to the low analysis acceptance for HSCPs produced with $\beta\approx 0.5$ and largely reconstructed in a single BX. }\label{fig:limits2}
\end{figure}

Upper limits at 95\% confidence level (\CL) are computed using the modified frequentist \CLs criterion~\cite{Junk:1999kv,Read:2002hq},
using the asymptotic approximation~\cite{Cowan:2010js}. 
The limits on the signal cross section are shown in Fig.~\ref{fig:limits}, as a function of the HSCP mass 
for the nonresonant $\PGtpr$ (\cmsLeft) and gluino $R$-hadron (\cmsRight) scenarios. 
The sensitivity is driven by 4-stub categories, in particular by the BX123 category, which features a 4-stub track across 
3 BXs. 
The 3-stub categories perform better for hadron-like models than for lepton-like ones, because the hadron-like HSCPs tend to be 
reconstructed with fewer stubs. 
The relative sensitivity of the
different categories is mass-dependent because heavier HSCPs typically have lower $\beta$ and take longer to cross the muon chambers. 
The structure around $m=3\TeV$ arises from the significant fraction of HSCPs with $\beta\approx0.5$, for which the analysis has 
limited acceptance since they tend to cross the muon detector entirely during the BX that follows the collision. 
Around $m=6\TeV$, the limits degrade because the HSCPs are mostly produced with $\beta\lesssim0.15$ and are stopped 
before crossing the muon detector. 
The previous CMS analysis based on an ionization-loss signature~\cite{CMS:2024nhn} provided an interpretation up to $m=2.6\TeV$, 
where the limits are about two orders of magnitude tighter than the result presented in this paper, because of the larger data set available, 
larger $\eta$ acceptance, and higher-level information. However, unlike the present search, 
the ionization-loss signature analysis does not have sensitivity to scenarios 
with $m\gtrsim4.5\TeV$ because its HLT efficiency is zero at such low $\beta$ values.
This analysis has unique sensitivity at high mass where all HSCPs have low $\beta$, and to the gluino $R$-hadron scenario with $f=1$, where there is no 
ionization loss signature in the tracker since the $R$-hadron is neutral in that part of the detector. 
Two-dimensional upper limits are shown in Fig.~\ref{fig:limits2} for the $\PGtpr$ pair production from a $\PZpr$ boson decay, with unique sensitivity for
$m_{\PZpr}\lesssim 2.5 m_{\PGtpr}$, corresponding to HSCPs with $\beta\lesssim0.6$. For all mass points, the observed limits are 
within the band containing 68\% of the limits expected under the background-only hypothesis. 

\begin{figure}[tb!]
  \centering
  \includegraphics[width=0.49\textwidth]{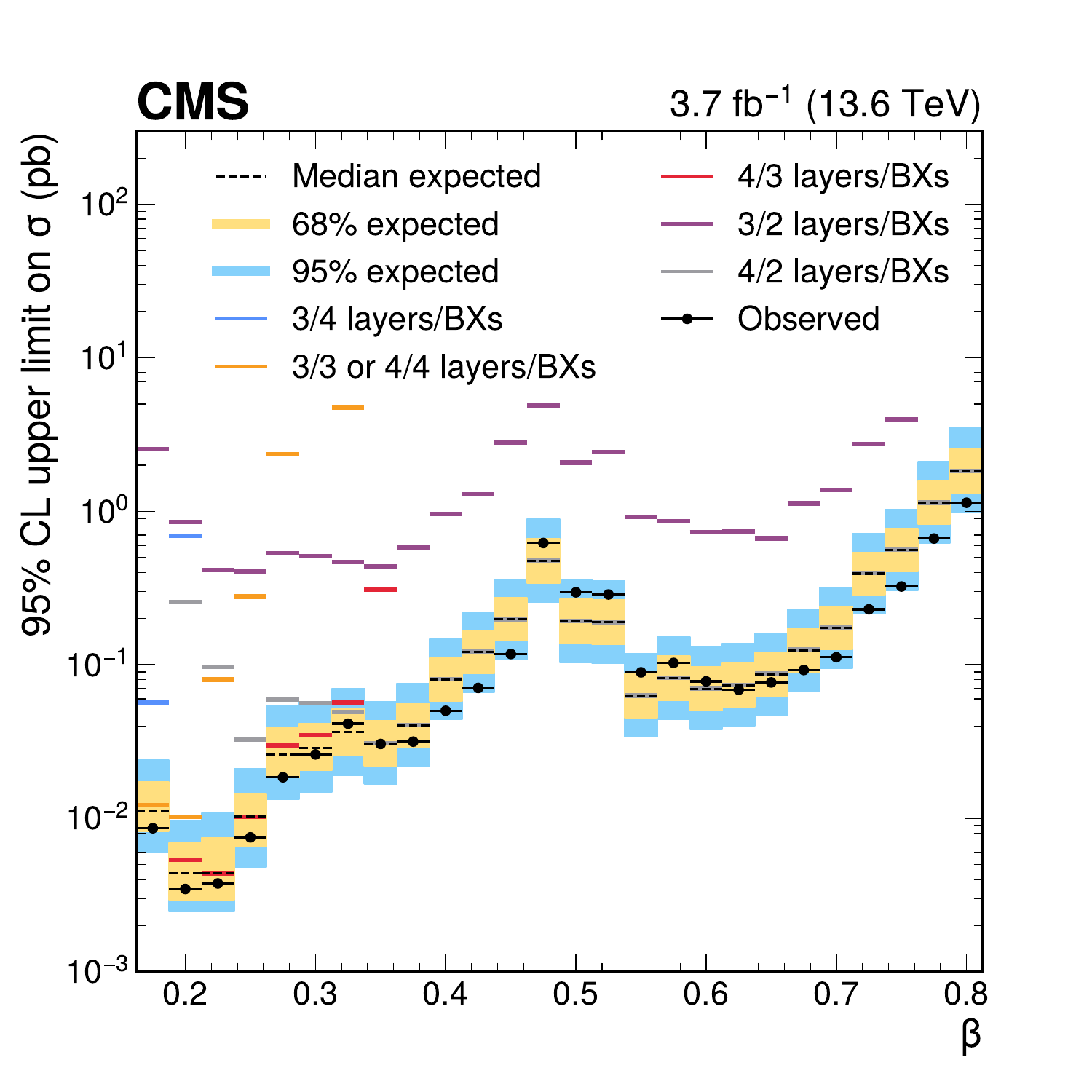}
	\caption{Observed and expected upper limits at 95\% \CL on the fiducial cross section of a heavy particle leaving a signature in the muon detector with $\abs{\eta}<0.83$, $\pt>500\GeV$, in bins of $\beta$.  The inner (yellow) and the outer (blue) bands indicate the regions containing 68 and 95\%, respectively, of the distribution of limits expected under the background-only hypothesis. The expected limits for groups of categories corresponding to a given number of muon detector layers across a given number of BXs are also shown with colored dashed lines. The limited sensitivity around $\beta=0.5$ corresponds to a loss of acceptance from particles reconstructed entirely in the BX that followed the collision.}\label{fig:limits_fid}
\end{figure}

Upper limits are also set on the fiducial cross section of a lepton-like particle leaving a signature 
in the muon detector and produced with $\pt>500\GeV$, $\abs{\eta}<0.83$, and a given range of $\beta$, 
assuming there is no contribution from particles produced with other $\beta$ values. 
These results are extracted from the SRs without additional track requirement in order not to rely 
on the pair production assumption. 
The signal distributions are built using particles from the simulated samples described in Section~\ref{sec:samples} 
satisfying $\pt>500\GeV$, $\abs{\eta}<0.83$, and the $\beta$ range restriction. These distributions are independent from the 
signal model. 
As shown in 
Fig.~\ref{fig:limits_fid}, upper limits as low as 3.5\unit{fb} can be set for particles produced with $0.1875<\beta<0.2125$, 
where the sensitivity mostly comes from the categories where the 
particle takes 3 BXs to cross all four muon layers. For lower $\beta$ values, the sensitivity sharply 
decreases as the particles get stopped before entering the muon 
detector because of their large ionization loss. The patterns in the limits align with the signatures of the slow particles as 
a function of $\beta$, shown in Fig.~\ref{fig:fraction}. For $\beta>0.35$, the categories where the particle takes 2 BXs to cross the four muon layers dominate the sensitivity. The sensitivity is low around $\beta=0.5$, because the particle is likely to 
be reconstructed entirely in the BX following the collision, where the analysis has no acceptance since the track 
experimentally looks like that of an ultra-relativistic particle. 

\section{Summary}

A search for long-lived massive charged particles has been presented, exploiting a muon-like signature in the barrel muon detectors
spread across several LHC proton bunch crossings.
This is the first time that an analysis relies on the novel CMS level-1 trigger scouting data set, for which no trigger selection is applied. 
A method based on control samples in data is used to estimate the backgrounds, 
relying on tracks not compatible in time with the expectations from a slow particle exiting the detector. No significant excess of data above the predicted standard model backgrounds is observed.   
Upper limits are set on the production cross section of heavy stable charged particles in several models, and fiducial upper 
limits on the production cross section for different $\beta=v/c$ ranges are also set for a model-independent interpretation. 
The analysis has unique 
sensitivity to particles with $0.15\lesssim \beta \lesssim 0.5$, which are challenging to trigger on, 
and to neutral particles that acquire a charge when crossing the
detector, which do not leave an ionization-loss signature in the tracker. Upper limits as low as 3.5\unit{fb} at 95\% confidence level 
are set on the fiducial production cross section 
of lepton-like charged particles with $\abs{\eta}<0.83$ and $\pt>500\GeV$, in several bins of $\beta$, extending the reach of the existing 
searches to higher masses and lower values of $\beta$. 

\begin{acknowledgments}
  We congratulate our colleagues in the CERN accelerator departments for the excellent performance of the LHC and thank the technical and administrative staffs at CERN and at other CMS institutes for their contributions to the success of the CMS effort. In addition, we gratefully acknowledge the computing centers and personnel of the Worldwide LHC Computing Grid and other centers for delivering so effectively the computing infrastructure essential to our analyses. Finally, we acknowledge the enduring support for the construction and operation of the LHC, the CMS detector, and the supporting computing infrastructure provided by the following funding agencies: SC (Armenia), BMBWF and FWF (Austria); FNRS and FWO (Belgium); CNPq, CAPES, FAPERJ, FAPERGS, and FAPESP (Brazil); MES and BNSF (Bulgaria); CERN; CAS, MoST, and NSFC (China); MINCIENCIAS (Colombia); MSES and CSF (Croatia); RIF (Cyprus); SENESCYT (Ecuador); ERC PRG, TARISTU24-TK10 and MoER TK202 (Estonia); Academy of Finland, MEC, and HIP (Finland); CEA and CNRS/IN2P3 (France); SRNSF (Georgia); BMFTR, DFG, and HGF (Germany); GSRI (Greece); NKFIH (Hungary); DAE and DST (India); IPM (Iran); SFI (Ireland); INFN (Italy); MSIT and NRF (Republic of Korea); MES (Latvia); LMTLT (Lithuania); MOE and UM (Malaysia); BUAP, CINVESTAV, CONACYT, LNS, SEP, and UASLP-FAI (Mexico); MOS (Montenegro); MBIE (New Zealand); PAEC (Pakistan); MES, NSC, and NAWA (Poland); FCT (Portugal);  MESTD (Serbia); MICIU/AEI and PCTI (Spain); MOSTR (Sri Lanka); Swiss Funding Agencies (Switzerland); MST (Taipei); MHESI (Thailand); TUBITAK and TENMAK (T\"{u}rkiye); NASU (Ukraine); STFC (United Kingdom); DOE and NSF (USA).

\hyphenation{Rachada-pisek} Individuals have received support from the Marie-Curie program and the European Research Council and Horizon 2020 Grant, contract Nos.\ 675440, 724704, 752730, 758316, 765710, 824093, 101115353, 101002207, 101001205, and COST Action CA16108 (European Union); the Leventis Foundation; the Alfred P.\ Sloan Foundation; the Alexander von Humboldt Foundation; the Science Committee, project no. 22rl-037 (Armenia); the Fonds pour la Formation \`a la Recherche dans l'Industrie et dans l'Agriculture (FRIA) and Fonds voor Wetenschappelijk Onderzoek contract No. 1228724N (Belgium); the Beijing Municipal Science \& Technology Commission, No. Z191100007219010, the Fundamental Research Funds for the Central Universities, the Ministry of Science and Technology of China under Grant No. 2023YFA1605804, the Natural Science Foundation of China under Grant No. 12535004, and USTC Research Funds of the Double First-Class Initiative No.\ YD2030002017 (China); the Ministry of Education, Youth and Sports (MEYS) of the Czech Republic; the Shota Rustaveli National Science Foundation, grant FR-22-985 (Georgia); the Deutsche Forschungsgemeinschaft (DFG), among others, under Germany's Excellence Strategy -- EXC 2121 ``Quantum Universe" -- 390833306, and under project number 400140256 - GRK2497; the Hellenic Foundation for Research and Innovation (HFRI), Project Number 2288 (Greece); the Hungarian Academy of Sciences, the New National Excellence Program - \'UNKP, the NKFIH research grants K 131991, K 133046, K 138136, K 143460, K 143477, K 146913, K 146914, K 147048, 2020-2.2.1-ED-2021-00181, TKP2021-NKTA-64, and 2025-1.1.5-NEMZ\_KI-2025-00004 (Hungary); the Council of Science and Industrial Research, India; ICSC -- National Research Center for High Performance Computing, Big Data and Quantum Computing, FAIR -- Future Artificial Intelligence Research, and CUP I53D23001070006 (Mission 4 Component 1), funded by the NextGenerationEU program (Italy); the Latvian Council of Science; the Ministry of Education and Science, project no. 2022/WK/14, and the National Science Center, contracts Opus 2021/41/B/ST2/01369, 2021/43/B/ST2/01552, 2023/49/B/ST2/03273, and the NAWA contract BPN/PPO/2021/1/00011 (Poland); the Funda\c{c}\~ao para a Ci\^encia e a Tecnologia, grant CEECIND/01334/2018 (Portugal); the National Priorities Research Program by Qatar National Research Fund; MICIU/AEI/10.13039/501100011033, ERDF/EU, ``European Union NextGenerationEU/PRTR", and Programa Severo Ochoa del Principado de Asturias (Spain); the Chulalongkorn Academic into Its 2nd Century Project Advancement Project, the National Science, Research and Innovation Fund program IND\_FF\_68\_369\_2300\_097, and the Program Management Unit for Human Resources \& Institutional Development, Research and Innovation, grant B39G680009 (Thailand); the Kavli Foundation; the Nvidia Corporation; the SuperMicro Corporation; the Welch Foundation, contract C-1845; and the Weston Havens Foundation (USA).
\end{acknowledgments}

\bibliography{auto_generated} 

\clearpage

\appendix

\section{Schematic illustration of the categories}
\label{app:schematic}

\begin{figure*}[htbp]
  \centering
        \includegraphics[width=0.45\textwidth]{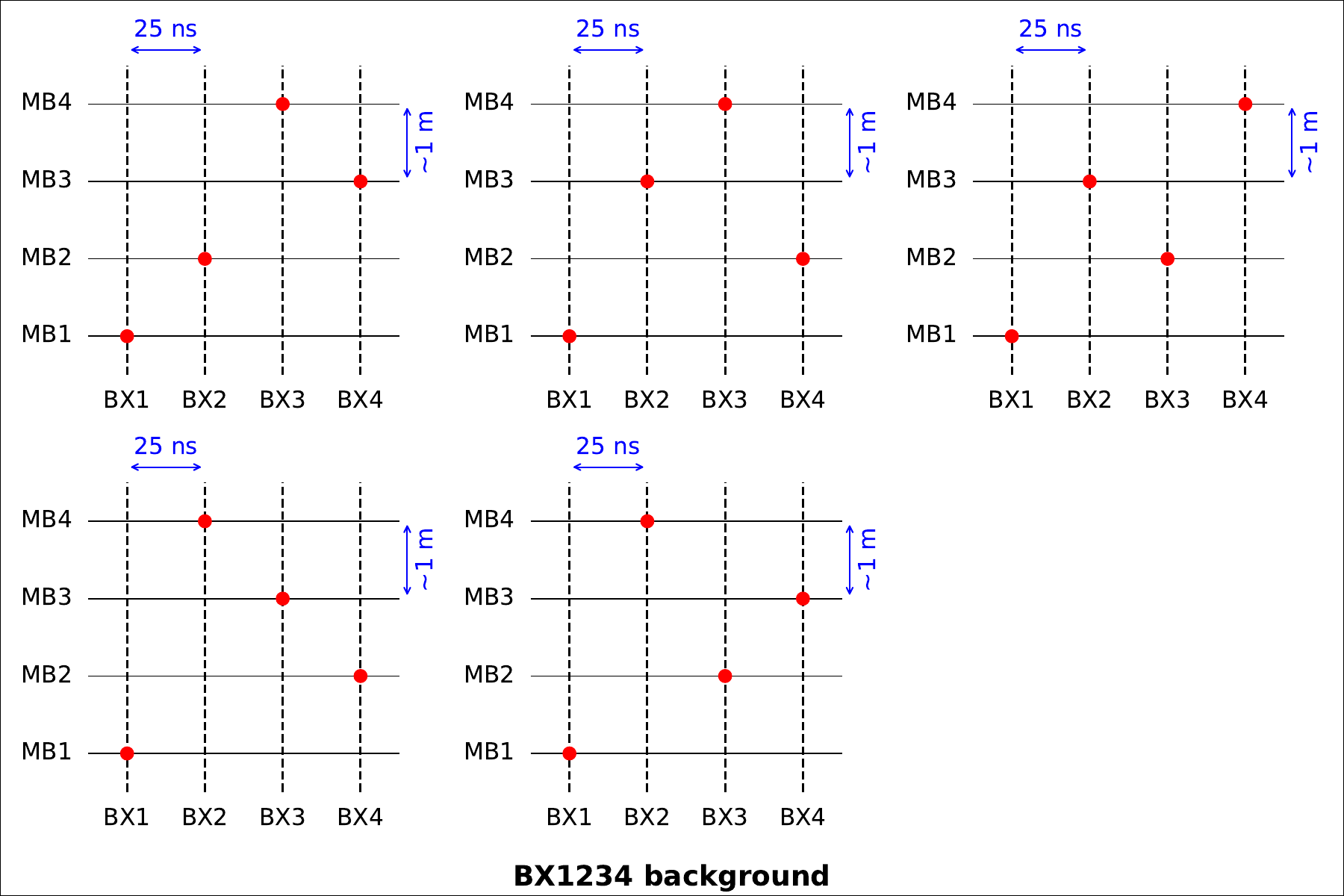}
        \includegraphics[width=0.45\textwidth]{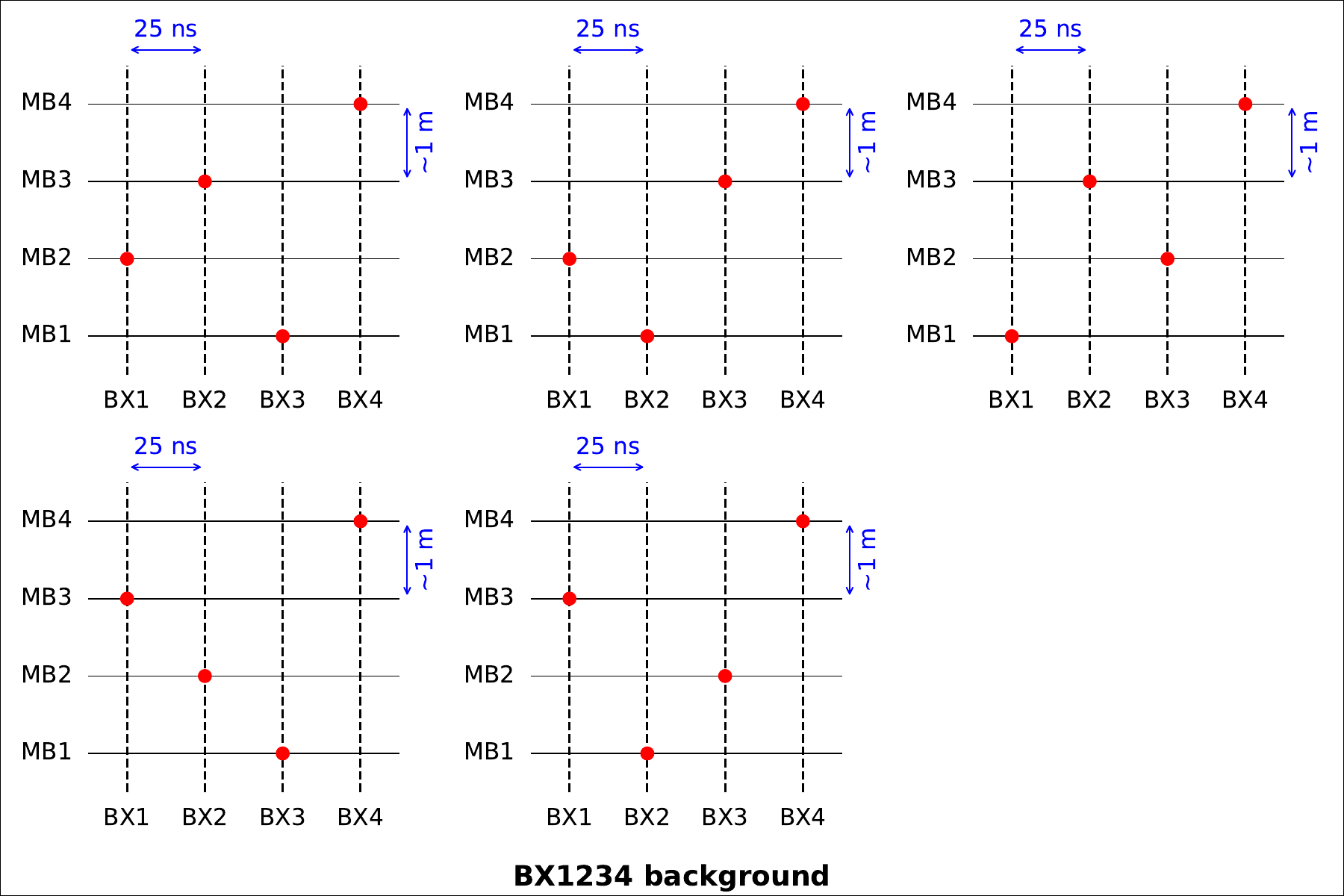}\\
        \includegraphics[width=0.25\textwidth]{Figure_014-c.pdf}
        \includegraphics[width=0.25\textwidth]{Figure_014-d.pdf}
        \includegraphics[width=0.25\textwidth]{Figure_014-e.pdf}\\
        \includegraphics[width=0.45\textwidth]{Figure_014-f.pdf}
        \includegraphics[width=0.45\textwidth]{Figure_014-g.pdf}
        \caption{Schematic definition of the categories with tracks across 2 BXs.
}\label{fig:cat_2BX}
\end{figure*}

\begin{figure*}[htbp]
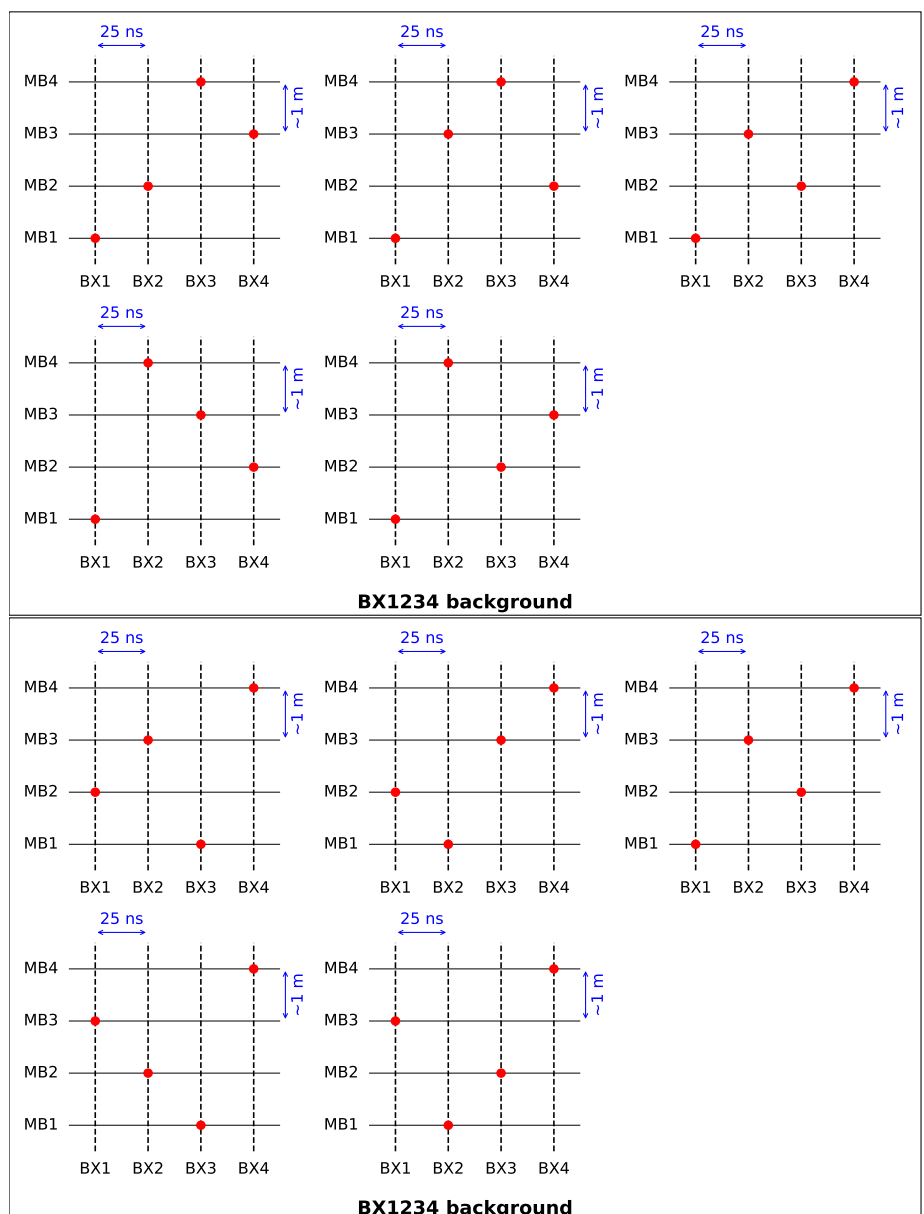

  \centering
        \includegraphics[width=0.75\textwidth]{Figure_015-a.pdf}
        \includegraphics[width=0.75\textwidth]{Figure_015-b.pdf}
        \caption{Schematic definition of the asynchronous orderings used to estimate the backgrounds in the BX1234 category. The upper (lower) half includes orderings with the first (last) stub in MB1 (MB4).
        }\label{fig:BX1234bkg}
\end{figure*}

\begin{figure*}[htbp]
  \centering
        \includegraphics[width=0.24\textwidth]{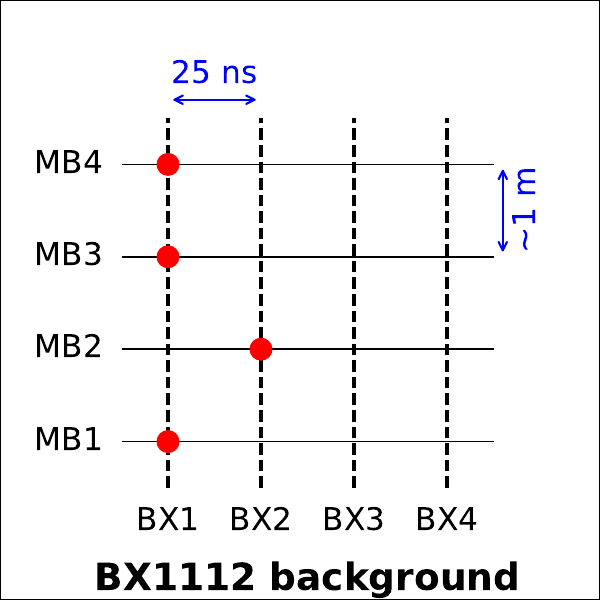}
        \includegraphics[width=0.24\textwidth]{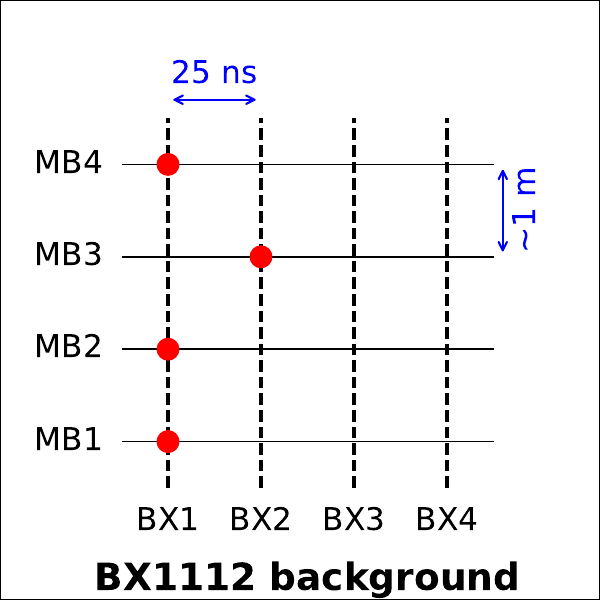}
        \caption{Schematic definition of the asynchronous orderings used to estimate the backgrounds in the BX1112 category. The left (right) diagram shows the ordering where the stub detected in a different BX than the other three stubs is in MB2 (MB3).
        }\label{fig:BX1112bkg}
\end{figure*}

The event categories are illustrated in Figs.~\ref{fig:cat_gt2BX_3stubs}--\ref{fig:cat_2BX}. The track 
configurations used to estimate the background in the BX1234 and BX1112 categories are illustrated in 
Figs.~\ref{fig:BX1234bkg} and \ref{fig:BX1112bkg}, respectively.
\begin{figure*}[ht!]
  \centering
        \includegraphics[width=0.49\textwidth]{Figure_012-a.pdf}
        \includegraphics[width=0.49\textwidth]{Figure_012-b.pdf}\\
        \includegraphics[width=0.49\textwidth]{Figure_012-c.pdf}
        \includegraphics[width=0.49\textwidth]{Figure_012-d.pdf}
        \caption{Schematic definition of the 3-stub categories with tracks across $>$2 BXs.
}\label{fig:cat_gt2BX_3stubs}
\end{figure*}

\begin{figure*}[htbp]
  \centering
        \includegraphics[width=0.75\textwidth]{Figure_013-a.pdf}\\
        \includegraphics[width=0.75\textwidth]{Figure_013-b.pdf}
        \includegraphics[width=0.25\textwidth]{Figure_013-c.pdf}\\
        \caption{Schematic definition of the 4-stub categories with tracks across $>$2 BXs.
}\label{fig:cat_gt2BX_4stubs}
\end{figure*}

\cleardoublepage \section{The CMS Collaboration \label{app:collab}}\begin{sloppypar}\hyphenpenalty=5000\widowpenalty=500\clubpenalty=5000
\cmsinstitute{Yerevan Physics Institute, Yerevan, Armenia}
{\tolerance=6000
A.~Hayrapetyan, V.~Makarenko\cmsorcid{0000-0002-8406-8605}, A.~Tumasyan\cmsAuthorMark{1}\cmsorcid{0009-0000-0684-6742}
\par}
\cmsinstitute{Institut f\"{u}r Hochenergiephysik, Vienna, Austria}
{\tolerance=6000
W.~Adam\cmsorcid{0000-0001-9099-4341}, L.~Benato\cmsorcid{0000-0001-5135-7489}, T.~Bergauer\cmsorcid{0000-0002-5786-0293}, M.~Dragicevic\cmsorcid{0000-0003-1967-6783}, P.S.~Hussain\cmsorcid{0000-0002-4825-5278}, M.~Jeitler\cmsAuthorMark{2}\cmsorcid{0000-0002-5141-9560}, N.~Krammer\cmsorcid{0000-0002-0548-0985}, A.~Li\cmsorcid{0000-0002-4547-116X}, D.~Liko\cmsorcid{0000-0002-3380-473X}, M.~Matthewman, J.~Schieck\cmsAuthorMark{2}\cmsorcid{0000-0002-1058-8093}, R.~Sch\"{o}fbeck\cmsAuthorMark{2}\cmsorcid{0000-0002-2332-8784}, M.~Shooshtari\cmsorcid{0009-0004-8882-4887}, M.~Sonawane\cmsorcid{0000-0003-0510-7010}, W.~Waltenberger\cmsorcid{0000-0002-6215-7228}, C.-E.~Wulz\cmsAuthorMark{2}\cmsorcid{0000-0001-9226-5812}
% ...
\par}
\vskip\cmsinstskip
\dag:~Deceased\\
$^{1}$Also at Yerevan State University, Yerevan, Armenia\\
$^{2}$Also at TU Wien, Vienna, Austria\\
% ...
\end{sloppypar}
%%% END EDITABLE REGION %%%
% skeleton_end
\end{document}